\documentclass[apj]{emulateapj}
\usepackage{amsmath}
\usepackage{graphicx}
\usepackage[hyperfootnotes=false,colorlinks,citecolor=blue,linkcolor=blue,urlcolor=blue]{hyperref}
\DeclareMathAlphabet{\mathbfit}{OT1}{cmr}{bx}{it}
\renewcommand{\vec}[1]{\mathbfit{#1}}
\newcommand{\mat}[1]{\mathbfit{#1}}
\newcommand{\Mo}{\mathrm{M}_{\odot}}

\begin{document}

\title{The impact of baryon physics on the structure of high-redshift galaxies}
\shorttitle{The impact of baryon physics on the structure of high-redshift galaxies}

\author{Marcel Zemp$^{1}$, Oleg Y. Gnedin$^{1}$, Nickolay Y. Gnedin$^{2,3,4}$ and Andrey V. Kravtsov$^{3,4}$}
\affil{$^{1}$ Department of Astronomy, University of Michigan, Ann Arbor, MI 48109, USA\\
$^{2}$ Particle Astrophysics Center, Fermi National Accelerator Laboratory, Batavia, IL 60510, USA\\
$^{3}$ Kavli Institute for Cosmological Physics, University of Chicago, Chicago, IL 60637, USA\\
$^{4}$ Department of Astronomy \& Astrophysics, University of Chicago, Chicago, IL 60637 USA}

\email{mzemp@umich.edu}

\shortauthors{Zemp et al.}

\begin{abstract}
We study the detailed structure of galaxies at redshifts $z \geq 2$ using cosmological simulations with improved modeling of the interstellar medium and star formation.
The simulations follow the formation and dissociation of molecular hydrogen, and include star formation only in cold molecular gas.
The molecular gas is more concentrated towards the center of galaxies than the atomic gas, and as a consequence, the resulting stellar distribution is very compact.
For halos with total mass above $10^{11}~\Mo$, the median half-mass radius of the stellar disks is 0.8 kpc at $z \approx 3$.
The vertical structure of the molecular disk is much thinner than that of the atomic neutral gas.
Relative to the non-radiative run, the inner regions of the dark matter halo change shape from prolate to mildly oblate and align with the stellar disk.
However, we do not find evidence for a significant ``dark disk'' of dark matter around the stellar disk.
The outer halo regions retain the orientation acquired during accretion and mergers, and are significantly misaligned with the inner regions.
The radial profile of the dark matter halo contracts in response to baryon dissipation, establishing an approximately isothermal profile throughout most of the halo.
This effect can be accurately described by a modified model of halo contraction.
The angular momentum of a fixed amount of inner dark matter is approximately conserved over time, while in the dissipationless case most of it is transferred outward during mergers.
The conservation of the dark matter angular momentum provides supporting evidence for the validity of the halo contraction model in a hierarchical galaxy formation process.
\end{abstract}


\section{Introduction}

With the ever increasing computer power and sophistication of algorithms, dissipationless cosmological simulations produced a consistent picture of the large-scale structure formation in the $\Lambda$CDM cosmology \cite[e.g.][]{2005Natur.435..629S, 2006ApJ...647..201C, 2009MNRAS.398.1150B, 2009A&A...497..335T} as well as the detailed structure of individual objects \cite[e.g.][]{2008Natur.454..735D, 2008MNRAS.391.1685S, 2009MNRAS.398L..21S}.
In these pure dark matter simulations, halos have a nearly universal density profile with a steep inner cusp \citep[e.g.][]{1991ApJ...378..496D, 1996ApJ...462..563N, 2010MNRAS.402...21N, 2005MNRAS.364..665D}, their overall shape is triaxial \citep[e.g.][]{1988ApJ...327..507F, 1991ApJ...368..325K, 2005ApJ...627..647B, 2006MNRAS.367.1781A, 2007MNRAS.376..215B}, and they exhibit a complex hierarchy of substructure \citep[e.g.][]{1999ApJ...524L..19M, 1999ApJ...522...82K, 2000ApJ...544..616G, 2008ApJ...680L..25D, 2009MNRAS.394..641Z, 2009MNRAS.395..797V, 2011MNRAS.413.1419V}.

Important missing ingredients in these dissipationless simulations are the physics of cosmic gas and the formation of stars.
In the large-scale structure simulations, it is an appropriate approximation to neglect cooling and star formation, since on these scales gravity is the only dynamically relevant force.
But on smaller scales it is necessary to include baryonic physics in order to resolve the internal structure of dark matter halos correctly since the condensation of baryons alters the phase-space structure of all matter components  \citep[e.g.][]{1994ApJ...431..617D, 2004ApJ...616...16G, 2004ApJ...611L..73K, 2006PhRvD..74l3522G, 2008ApJ...681.1076D,2010MNRAS.402..776P,2010MNRAS.405.1119K,2010MNRAS.405.2161D,2010MNRAS.406..922T,2010MNRAS.407..435A}.

The processes of star formation and its feedback on the interstellar medium are complex in nature and not well understood.
Hence, often crude, empirical prescriptions have been used in order to model these processes in simulations.
A common example is a Kennicutt-Schmidt-type recipe based on the total gas mass density $\rho_\mathrm{G}$ on small scales of the form $\dot{\rho}_\mathrm{S} \propto \rho_\mathrm{G}^{3/2}$.
Fine tuning is then achieved with the help of threshold criteria (e.g. for temperature and/or density) in order to reproduce the observed Kennicutt-Schmidt relation,
$\dot{\Sigma}_\mathrm{S} \propto \Sigma_\mathrm{G}^{1.4}$ \citep{1959ApJ...129..243S, 1998ApJ...498..541K}, where $\dot{\Sigma}_\mathrm{S}$ is the star formation rate surface density and $\Sigma_\mathrm{G}$ is the total gas surface density (see \citealt{2008MNRAS.383.1210S} and \citealt{2008ASL.....1....7M} for recent reviews).

\begin{deluxetable*}{ccccc}[t]
	\tablecolumns{5}
	\tablewidth{0pt}
  \tablecaption{\label{tab:simulationsummary}}
	\tablehead{\colhead{Simulation} & \colhead{Non-equilibrium} & \colhead{Star} & \colhead{Supernova} & \colhead{Supernova}\\
	 & \colhead{cooling} & \colhead{formation} & \colhead{metal enrichment} & \colhead{thermal feedback}}
	\startdata
	A & yes & yes & yes & yes \\
	A$_\mathrm{NF}$ & yes & yes & yes & no  \\
	B & no  & no  & no  & no
	\enddata
\end{deluxetable*}

Recent observations indicate that the Kennicutt-Schmidt relation shows more complexity that cannot be captured by a single power law over a wide range of total gas surface densities \citep{2008AJ....136.2782L, 2008AJ....136.2846B} and that the star formation rate surface density correlates better with the surface density of molecular hydrogen, $\Sigma_\mathrm{H_2}$ \citep{2002ApJ...569..157W, 2010MNRAS.407.2091G, 2011ApJ...730L..13B}.
The complex behavior of the Kennicutt-Schmidt relation as a function of the total gas surface density can be understood if star formation only happens in regions where the gas is predominantly in the molecular rather than atomic phase \citep{2003ApJ...590L...1K, 2006ApJ...645.1024P, 2008ApJ...680.1083R, 2009ApJ...707..954P, 2009ApJ...699..850K, 2009ApJ...697...55G, 2010ApJ...714..287G}.
While this is still a working hypothesis, theoretical studies demonstrated that the transition from the atomic to molecular gas may also correspond to the conditions under which gas becomes susceptible to gravitational fragmentation \citep{2011ApJ...731...25K, 2011arXiv1105.3073G}.
Hence, in order to realistically model star formation in simulations, it is necessary to follow the formation and dissociation of molecular hydrogen in the galactic interstellar medium.

In this paper we study the formation of high-redshift galaxies, using new state-of-the-art modeling of galactic interstellar medium and star formation.
We use the phenomenological molecular hydrogen modeling introduced in \cite{2011ApJ...728...88G} and a star formation recipe based on the density of molecular hydrogen.
With these simulations we explore the impact of baryons on the structure of the dark matter halos, important for many current observational studies.
We investigate the shape of dark matter halos as a function of radius, important for gravitational lensing studies; the response of the radial halo profile to central condensation of baryons, for studies of the Tully-Fisher and other galactic scaling relations; the alignment of dark halos and stellar disks; and the distribution and evolution of the angular momentum of dark matter.

\section{Simulations}

\subsection{Initial conditions}

We run cosmological simulations of a periodic box with comoving length $L_\mathrm{box} = 25.6\, h^{-1}\, \mathrm{Mpc}$.
We adopt a $\Lambda$CDM cosmology with a total matter density parameter $\Omega_{\mathrm{M},0} = 0.28$, dark matter density $\Omega_{\mathrm{DM},0} = 0.234$, baryon density $\Omega_{\mathrm{B},0} = 0.046$, cosmological constant $\Omega_{\Lambda} = 0.72$, Hubble parameter $H_0 = 100 h\, \mathrm{km}\, \mathrm{s}^{-1}\, \mathrm{Mpc}^{-1}$, with $h=0.7$, linearly extrapolated normalization of the power spectrum $\sigma_8 = 0.82$, and spectral index $n_s = 0.96$, consistent with the Wilkinson Microwave Anisotropy Probe 7-year data \citep{2011ApJS..192...14J}.

The initial conditions were generated taking into account a non-zero DC mode \citep{2005ApJ...634..728S,2011ApJS..194...46G}.
The DC mode corrects for a possible deviation of the average matter density in the box from the universal value, arising from the finite simulation volume.
It is equivalent to having a constant overdensity at redshift $z=0$ of $\delta_\mathrm{DC,0} \equiv \bar{\rho}_\mathrm{box,0}/\bar{\rho}_\mathrm{uni,0} - 1$.
In general, the DC mode denotes an offset of the mean of a signal or waveform from zero.
Usually, it is common practice to ignore the DC mode, but this is already a constraint on the initial conditions and therefore the initial conditions are not a truly random realization.
The positions and velocities of particles are determined by the usual Zel'dovich approximation \citep{1970A&A.....5...84Z, 1983MNRAS.204..891K, 1985ApJS...57..241E}.

First, we ran 5 low resolution (256$^3$ particles) random realizations of the cosmological box.
In these low-resolution simulations, we only simulate dissipationless evolution with an N-body code PKDGRAV2 \citep{2001PhDT........21S}.
We then selected a representative box that had the mass function of halos at redshift $z=0$ closest to that expected for the whole universe \citep[e.g.,][]{1999MNRAS.308..119S}.

This box has a DC mode of $\delta_\mathrm{DC.0} = 0.571$.
In this selected box we chose to refine 7 objects in the mass range $M_\mathrm{200b} \approx 10^{11} - 10^{13}~\Mo$ at $z=0$ (see also Figure~\ref{fig:halomass}).
$M_\mathrm{200b}$ is the mass within $r_\mathrm{200b}$ such that the enclosed density is $200 \rho_\mathrm{b}$, with $\rho_\mathrm{b}$ being the background matter density.
The selected objects had a quiet merger history after $z \approx 2$ but apart from that, they were selected randomly by visual examination.
We used the traditional method of refining a region of interest with a large number of dark matter particles and leaving the rest at lower resolution so that we correctly account for the large-scale tidal forces \citep{1991ApJ...368..325K,2001ApJS..137....1B}.
Around the 7 selected objects we refined a region of $5~r_\mathrm{200b}$ with an effective dark matter resolution of $2048^3$, i.e. the high resolution dark matter particles have a mass of $1.81 \times 10^5~\Mo$.
In the surrounding region, we increased the dark matter mass in buffer zones by factors of $2^3=8$ until we reached the initial low resolution level.
For the thickness of the buffer zones we chose 3 lengths of the top-level cells ($3 L_0$), where $L_0 = L_\mathrm{box}/256$ = 143 kpc (comoving).

The gas was initialized following the same refinement pattern with the baryonic power spectrum.
The initial composition of the gas is primordial, i.e. a hydrogen mass fraction $X=0.76$, helium mass fraction $Y = 0.24$, and metal mass fraction $Z=0$.
Further, we set $X_\mathrm{H\,\textsc{ii}} = 1.2 \times 10^{-5} \sqrt{\Omega_\mathrm{M,0}}/(\Omega_\mathrm{B,0} h) ~X = 1.5 \times 10^{-4}$ \cite[equation 6-119]{1993ppc..book.....P}, $X_\mathrm{H_2} = 2 \times 10^{-6} ~X = 1.52 \times 10^{-6}$ \citep{2001ApJ...560..580R}, and $X_\mathrm{H\,\textsc{i}} = X - X_\mathrm{H\,\textsc{ii}} - X_\mathrm{H_2}$.
All the helium is initially in the form of He\,\textsc{i}, i.e. $Y_\mathrm{He\,\textsc{i}} = Y$, $Y_\mathrm{He\,\textsc{ii}} = 0$, $Y_\mathrm{He\,\textsc{iii}} = 0$.

The starting redshift of the refined initial conditions was determined by the criterion that the root mean square of the density fluctuations is 0.1, which resulted in $z_\mathrm{IC} = 108$.

\begin{figure*}
	\centering
	\includegraphics[width=0.49\textwidth]{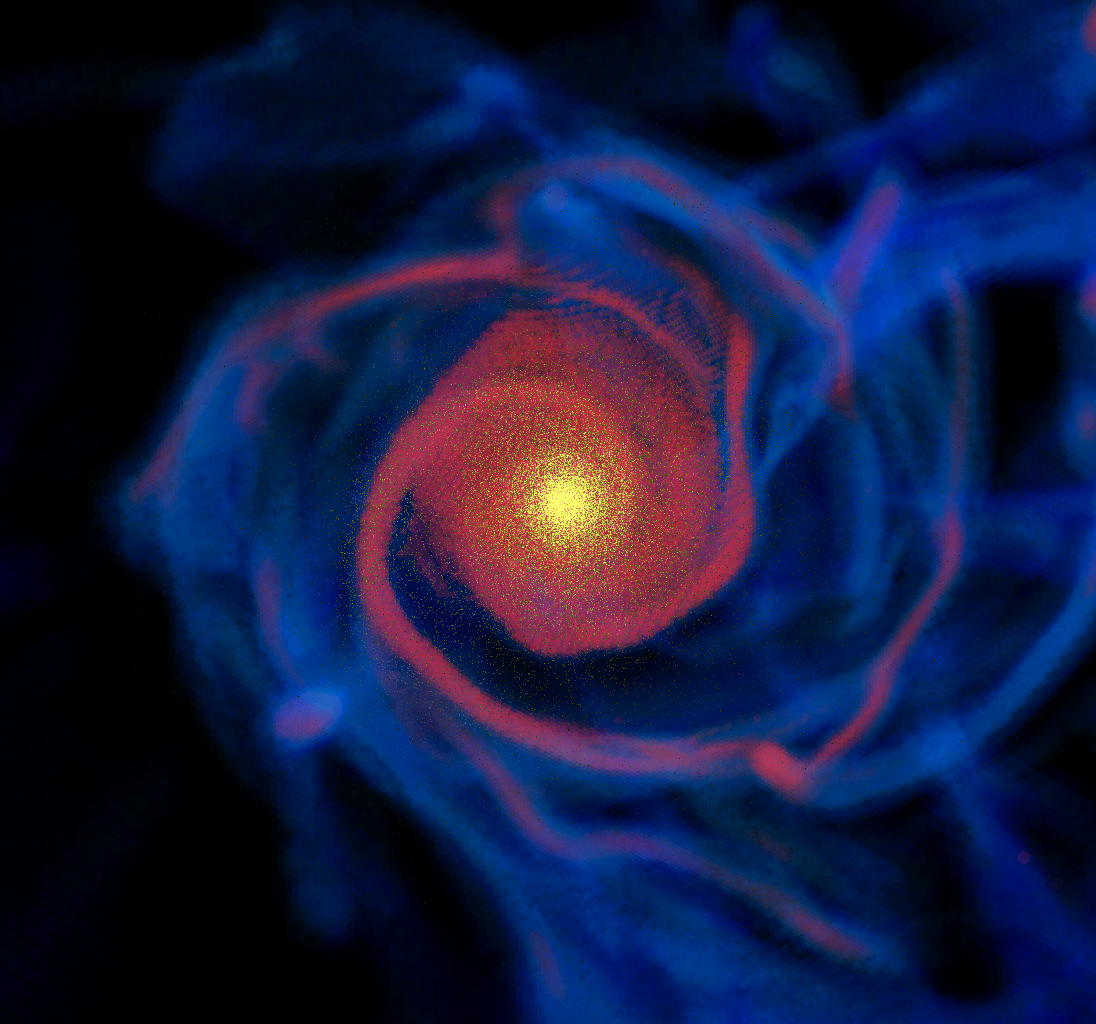}
	\includegraphics[width=0.49\textwidth]{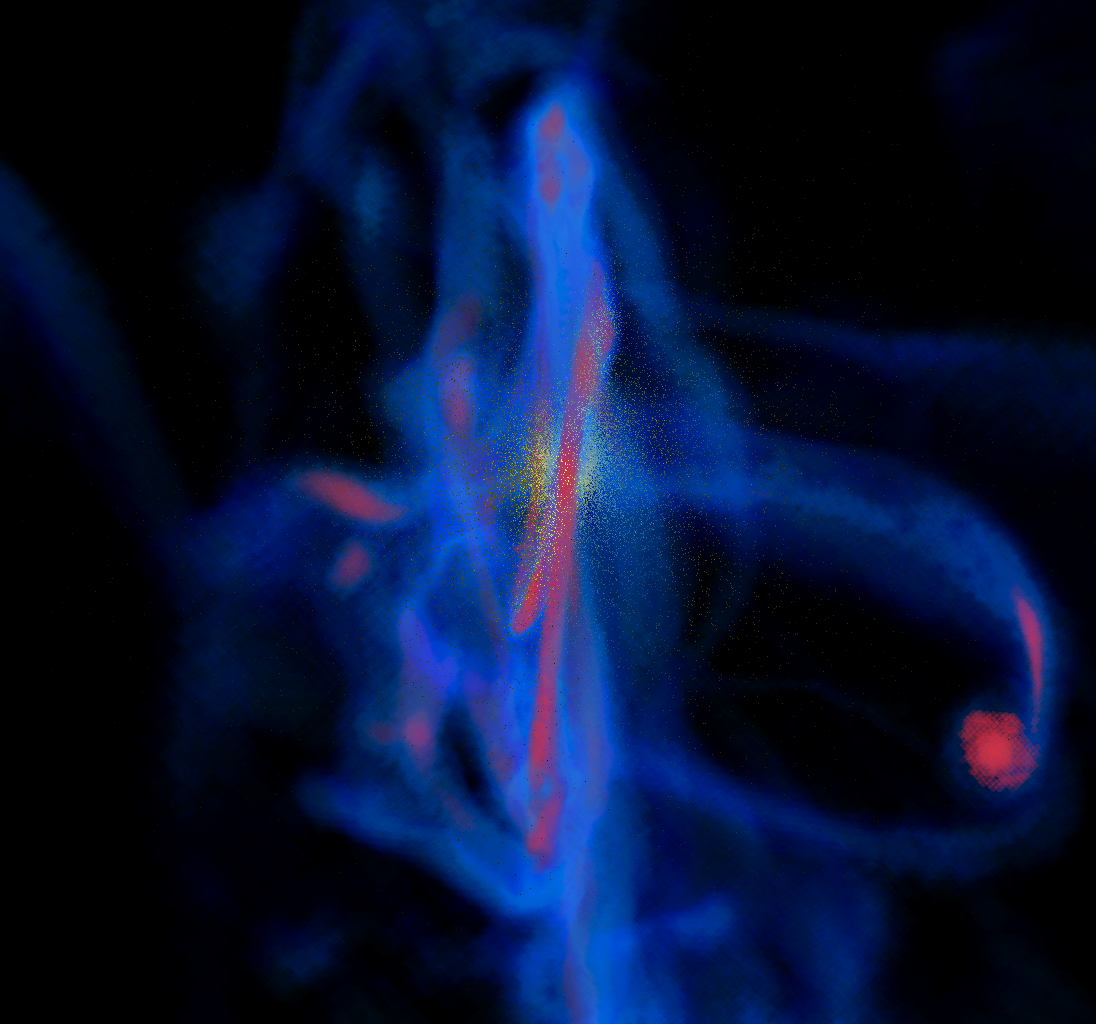}
	\caption{The disk of the most massive galaxy at $z \approx 3$ (the main halo) in simulation A in face-on (left panel) and edge-on (right panel) view.
	The molecular hydrogen is shown in red, atomic gas (H\,\textsc{i} and He\,\textsc{i}) in blue.
	The stars (yellow points) only form in the region where the molecular gas resides.
	The linear size of the image is $\approx 14$~kpc (physical).}
	\label{fig:disc}
\end{figure*}

\subsection{Input physics}\label{sec:simulations}

The simulations were run with the latest version of the gas dynamics and $N$-body Adaptive Refinement Tree (ART) code \citep{1997ApJS..111...73K,1999PhDT........25K,2002ApJ...571..563K,2008ApJ...672...19R}.
The Poisson and fluid equations are solved in super-comoving coordinates using cell-based adaptive mesh refinement (AMR) techniques \citep{1997ApJS..111...73K,2002ApJ...571..563K}.
ART includes 3-dimensional radiative transfer of ultraviolet (UV) radiation from individual stellar particles and from the extragalactic background (modeled according to \citealt{2001cghr.confE..64H}) using the Optically Thin Variable Eddington Tensor approximation \citep{2001NewA....6..437G}.
It includes a non-equilibrium chemical network for hydrogen (H\,\textsc{i}, H\,\textsc{ii}, and H$_2$) and helium (He\,\textsc{i}, He\,\textsc{ii}, and He\,\textsc{iii}) as well as non-equilibrium cooling and heating rates, which use the local abundances of atomic, molecular, and ionic species and the local UV intensity \citep{2011ApJ...728...88G}.
All these reactions are followed self-consistently in the course of a simulation.
An empirical model for the formation and shielding of molecular hydrogen on the interstellar dust allows for more realistic star formation recipes based on the local density of molecular hydrogen \citep{2011ApJ...728...88G}.
ART also includes metal enrichment and thermal feedback due to the Type II and Type Ia supernovae \citep{2003ApJ...590L...1K}.

The local star formation rate volume density $\dot{\rho}_\mathrm{S}$ in a cell is calculated as
\begin{equation}\label{eq:sfr}
	\dot{\rho}_\mathrm{S} = \epsilon_\mathrm{ff} \frac{\rho_\mathrm{H_2}}{\tau_\mathrm{sf}}
\end{equation}
where $\rho_\mathrm{H_2}$ is the local mass density of molecular hydrogen.
The star formation time scale is given by
\begin{equation}
	\tau_\mathrm{sf} = \mathrm{min}[\tau_\mathrm{ff}(\rho_\mathrm{G}),\tau_\mathrm{max}]
\end{equation}
where $\tau_\mathrm{ff} = (32 G \rho_\mathrm{G} / 3\pi)^{-1/2}$ is the free-fall time, and $\rho_\mathrm{G}$ the local gas density (including all hydrogen and helium species).
The maximum timescale $\tau_\mathrm{max}$ is set to the free-fall time of gas with a hydrogen number density of $n_\mathrm{H} = n_\mathrm{H\,\textsc{i}}+n_\mathrm{H\,\textsc{ii}}+2 n_\mathrm{H_2} = 50 ~\mathrm{cm}^{-3}$.
The star formation efficiency per local free-fall time is set to $\epsilon_\mathrm{ff} = 0.007$.
To ensure that star formation happens only in our numerical analogs of real molecular clouds, we allow star formation only in cells with the molecular mass fraction above $f_\mathrm{H_2} = 2 n_\mathrm{H_2}/n_\mathrm{H} = 0.1$.
These cells have a range of total gas density from 50 to $10^4$ amu cm$^{-3}$ for the  main halo at $z \approx 3$ (Figure \ref{fig:disc}).
Stellar particles are created via a Poisson process with a characteristic timescale of $2 \times 10^{7}$ yr.
This star formation prescription is similar to the recipe SF2 in \cite{2009ApJ...697...55G}.

Figure~\ref{fig:disc} shows the disc of the most massive galaxy in our simulation at $z \approx 3$, which we call the main halo.
The molecular hydrogen forms only in high density regions and hence the stars are confined to these central regions.
In traditional star formation prescriptions based on the total gas density instead of the molecular hydrogen density, stars would be formed over a much larger volume filled with the lower-density atomic gas.

We ran three versions of the simulation, which are summarized in Table~\ref{tab:simulationsummary}.
Simulation A is a full physics run with radiative transfer and non-equilibrium cooling.
Simulation A$_\mathrm{NF}$ is the same as simulation A but without supernova thermal feedback.
Metal enrichment due to supernovae is still included.
Simulation B is a non-radiative version without cooling and star formation.

In all simulations, the top level $l=0$ grid is $256^3$ and we allow for up to 9 more refinement levels ($l_\mathrm{max}=9$) where each higher level is refined by a factor 2 with respect to the parent level.
This results in a size of the smallest cells $L_9 = L_\mathrm{box}/(256 \cdot 2^9) = 279 ~\mathrm{pc}$ (comoving).
A cell is refined if its dark matter or gas mass exceeded $1.07 \times 10^{6} ~\Mo$ or $1.33 \times 10^{5} ~\Mo$, respectively. 
For the dark matter, this threshold corresponds to the mass of about 6 high resolution particles.
On each refinement level $l$, the time step is refined as well according to $\Delta \nu_l = \Delta \nu_0 / 2^l$, where $\Delta \nu_0$ is the global time step on the top level mesh.
The value of $\Delta \nu_0$ is set at the beginning of each top level step so that the Courant-Friedrichs-Lewy condition \citep{1928MatAn.100...32C,1967IBMJ...11..215C} is fulfilled on all levels \citep{2002ApJ...571..563K}.
In total our simulation A contains $2.89 \times 10^8$ dark matter particles and $3.89 \times 10^8$ gas cells at $z \approx 2$.

\subsection{Halo selection}\label{sec:haloselection}

\begin{figure}
	\centering
	\includegraphics[width=\columnwidth]{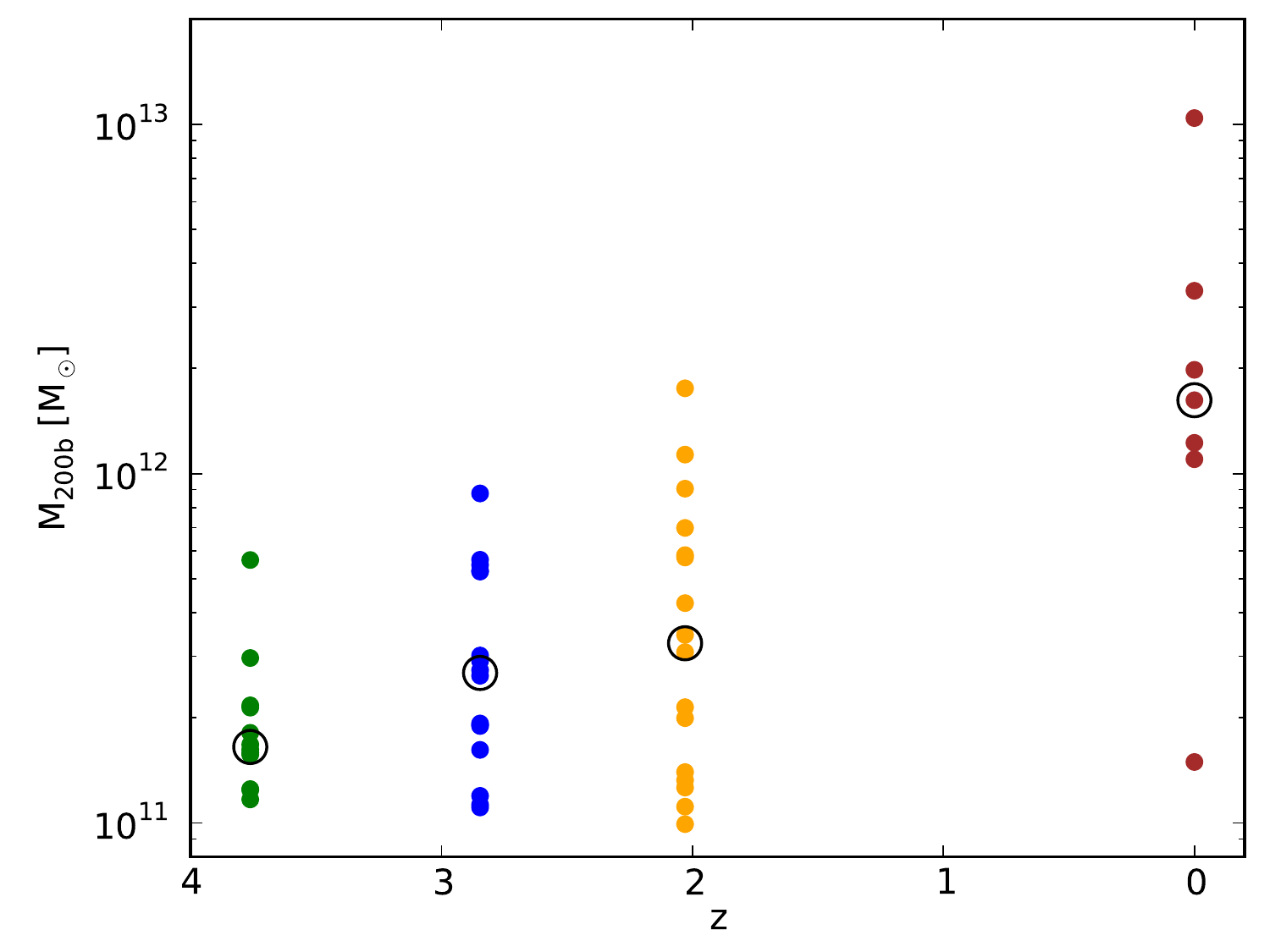}
	\caption{Masses of the selected halos at redshifts $z \approx$ 4, 3 and 2 in simulation A, as well as the final host halos at $z=0$ from the N-body simulation.
	The median mass at each epoch is indicated by a circle.
	The lower selection cutoff is a fixed mass of $10^{11}\, \Mo$, which results in different number of objects at different epochs.}
	\label{fig:halomass}
\end{figure}

For the analysis presented in this paper, we mainly concentrate on three snapshots at redshifts around 4, 3, and 2 (the exact redshifts are 3.76, 2.85, and 2.03) in run A.
These epochs correspond to 1.69 Gyr, 2.32 Gyr, and 3.29 Gyr after the Big Bang in our cosmology.
The output redshifts for different runs match to within $\Delta z = 0.005$.

Simulation A$_\mathrm{NF}$ at full resolution was stopped at $z = 2.77$ to save computing time.
We ran also a lower-resolution version of A$_\mathrm{NF}$ with the 8 times more massive dark matter particles but otherwise the same gas physics and parameters.
We use the $z \approx 2$ snapshot from this version in our analysis.

In all snapshots we ran a variant of the Bound Density Maxima halo finder \citep{2004ApJ...609..482K} and selected all massive objects with $M_\mathrm{200b} \geq 10^{11}\, \Mo$ in the high resolution region of run A.
We then matched simulation A with the other two simulations in order to find the corresponding halos in these runs.
The matching procedure is not a trivial task since the dynamics in the various runs is different and the output epochs do not match exactly (e.g. halos already could have merged at a given time).
A simple matching scheme based on comoving position and mass is not always successful, though it gives a reasonable initial guess.
We therefore check all the matches by eye and repair possible misidentifications.
This procedure is rather cumbersome but absolutely necessary in order to get quantitatively reliable results.
We only use halos where we could clearly identify a match in all three simulations.
There are 12, 16, and 16 objects (from a total of 16, 22, and 26 objects with $M_\mathrm{200b} \geq 10^{11}\, \Mo$) at redshifts $\approx$ 4, 3, and 2, respectively, that fulfill our selection and quality criteria (see Figure~\ref{fig:halomass}).

\subsection{Spatial resolution}

The force resolution in ART is roughly 2 cell sizes \citep{1997ApJS..111...73K} but numerical relaxation processes could affect the structure of halos on even larger scales.
In order to estimate the scale where numerical effects become important, we use the local relaxation time at radius $r$
\begin{equation}\label{eq:trelax}
	t_\mathrm{rel}(r) \equiv \frac{N(r)}{\ln[N(r)]} ~ t_\mathrm{dyn}(r)
\end{equation}
with
\begin{equation}
  t_\mathrm{dyn}(r) \equiv 2 \pi \sqrt{\frac{r^3}{G M(r)}}
\end{equation}
where $N(r)$ is the enclosed number of collisionless particles (dark matter and stars), and the dynamical time is set by the total enclosed matter (including gas), $M(r)$.
Here we use a slightly different normalization than in the usual expression for the local relaxation time \cite[e.g.,][]{2003MNRAS.338...14P} since we dropped the factor of 8 in front of the logarithmic term in the denominator.
\cite{2008MNRAS.386.1543Z} found that this normalization agrees better with the results of N-body simulations, where they studied the stability of isolated high-resolution halos.
Relaxation processes become important on the scale $r_\mathrm{rel}$ at time $t$ given by the solution of $t_\mathrm{rel}(r_\mathrm{rel}) = t$.

We checked for all our selected objects that $r_\mathrm{rel} \approx 2-4 ~L_9$.
Generally, dissipative runs A and A$_\mathrm{NF}$ have a smaller relaxation scale than run B, due to the higher number of particles within a given radius.
Therefore, we estimate that our results are numerically converged on scales larger than $r_\mathrm{res} = 4 ~L_9 = 1.12 ~\mathrm{kpc}$ (comoving).
We mark the resolution scale $r_\mathrm{res}$ in all the plots where appropriate.

\subsection{Median properties}

In the following sections, we present halo properties that were calculated using a profiling routine described in the Appendix.
We calculate the median value of each property for all selected halos as a function of radius.
In order to display the spread among individual halos, we also calculate the region between the 15th and 85th percentile values, i.e. the region containing 70 percent of the halos.
This is the spread specification throughout the paper, unless otherwise noted.

The median values are a useful representation because the selected objects are all similar, with the mass span of at most of an order of magnitude.
The most massive halo in run A at the three epochs (which is not the same halo) has $M_\mathrm{200b} = 5.66 \times 10^{11}~\Mo$, $8.77 \times 10^{11}~\Mo$, and $1.75 \times 10^{12}~\Mo$ at $z \approx 4$, 3, and 2, respectively (Figure~\ref{fig:halomass}).
The median mass and the 15th and 85th percentiles of the 12, 16 and 16 objects at these epochs are $M_\mathrm{200b} = 1.7^{+0.8}_{-0.4}\times 10^{11}\, \Mo$, $2.7^{+2.7}_{-1.3}\times 10^{11}\, \Mo$, and $3.3^{+5.3}_{-2.0}\times 10^{11}\, \Mo$, respectively.
The median virial radius and the percentiles are $r_\mathrm{200b} = 35^{+5}_{-3}$ kpc, $50^{+13}_{-11}$ kpc, and $67^{+25}_{-18}$ kpc, respectively.
The final masses at $z=0$ of the 7 host halos are $M_\mathrm{200b} = 1.6^{+2.4}_{-0.6}\times 10^{12}\, \Mo$.

\section{Spherically averaged properties}

\subsection{Density profile}

\begin{figure}[t]
	\centering
	\includegraphics[width=\columnwidth]{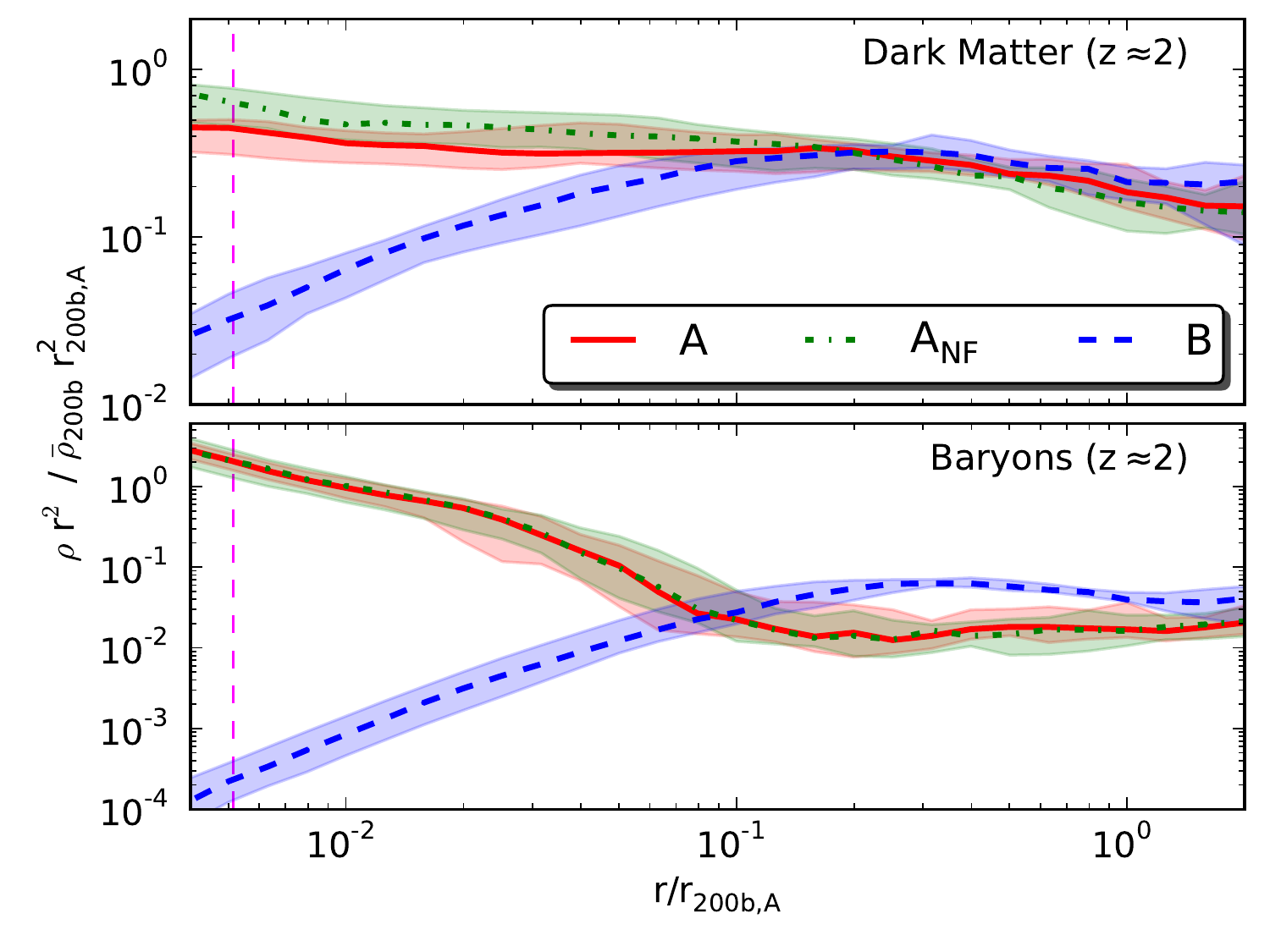}
	\caption{Median local mass density as a function of radius for the dark matter (top panel) and baryons (bottom panel) at $z \approx 2$.
	We plot the quantity $\rho r^2$ in order to reduce the dynamical range.
	The shaded areas mark the region containing 70 percent of the halos.
	The median resolution scale is marked with a vertical dashed line.
	The three simulations are described in Table~\ref{tab:simulationsummary}.}
	\label{fig:density}
\end{figure}

\begin{figure}[t]
	\centering
	\includegraphics[width=\columnwidth]{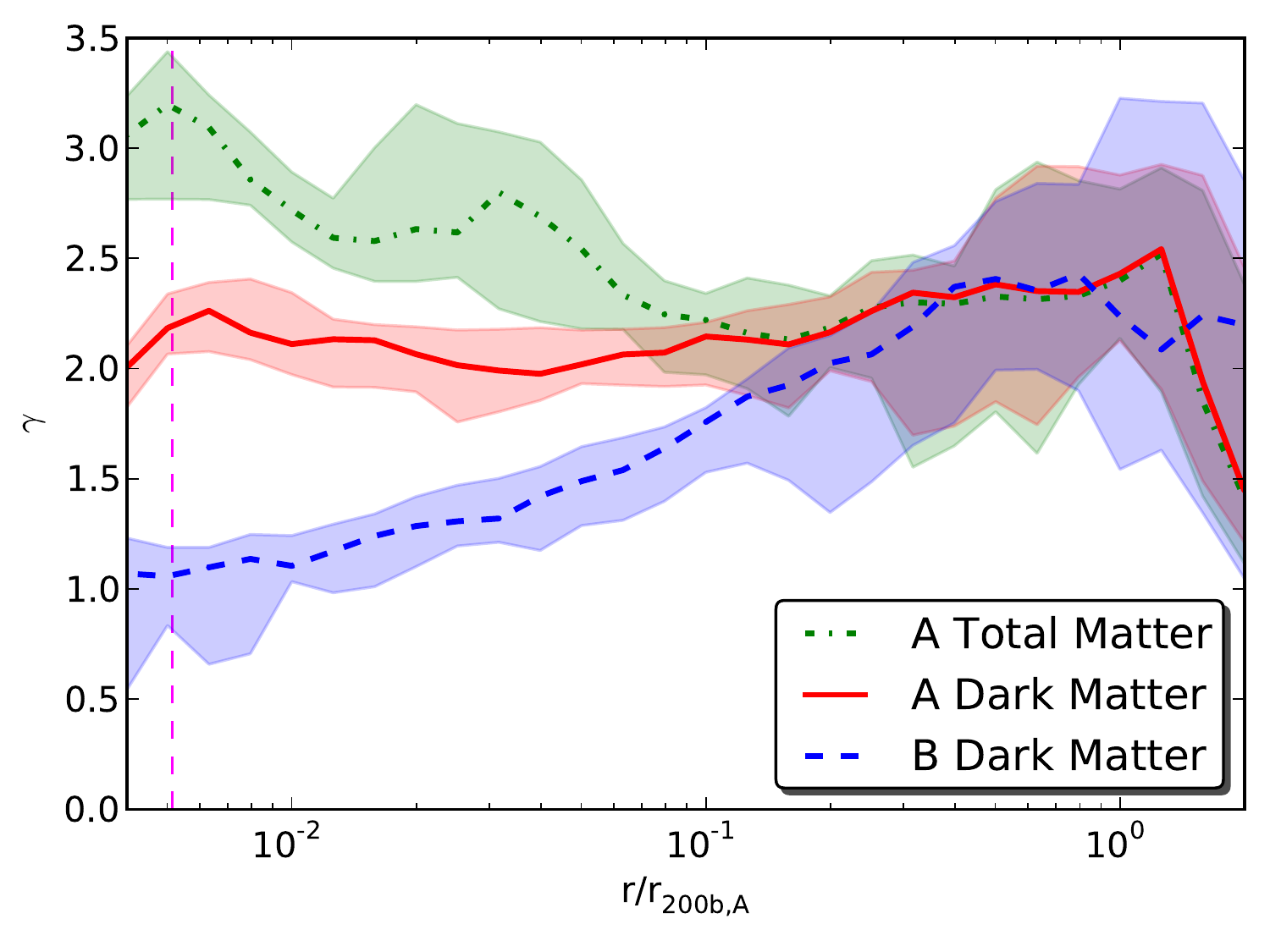}
	\caption{Median local mass density slope as a function of radius for the dark matter in simulations A and B as well as for the total matter in simulation A, at $z \approx 2$.
	The shading is as in Figure~\ref{fig:density}.}
	\label{fig:slope}
\end{figure}

In order to get a first impression of the impact of baryon physics on the overall matter distribution, we show in Figure~\ref{fig:density} the local mass density $\rho$ and in Figure~\ref{fig:slope} the local density slope $\gamma(r) \equiv - \mathrm{d}\log{\rho}/\mathrm{d}\log{r}$.
We normalize the radii of each halo by $r_\mathrm{200b,A}$, the virial radius of this halo in run A.
With this choice we always compare the same \textit{physical} scale of a specific halo in all three different simulations and avoid complications arising from the slightly different values of $r_\mathrm{200b}$ in the three runs.
The densities are normalized by $\bar{\rho}_\mathrm{200b} \equiv 3 M_\mathrm{200b}/4 \pi r_\mathrm{200b}^3 = 200 \rho_\mathrm{b}$.
In order to reduce the dynamical range, we plot the quantity $\rho r^2$ in Figure~\ref{fig:density}.

The most obvious difference is in the distribution of the baryons.
While in the non-radiative case the baryons form an approximately constant density core, they form a steep density profile with $\gamma \approx 3.5$ in the center for the cases A and A$_\mathrm{NF}$.
The central region is dominated by the stars.

In the non-radiative case B, the dark matter density profile is roughly proportional to $r^{-1}$ near the center, in agreement with other dissipationless simulations \citep[e.g.][]{1991ApJ...378..496D, 1996ApJ...462..563N, 2005MNRAS.364..665D, 2008Natur.454..735D, 2009MNRAS.398L..21S, 2010MNRAS.402...21N}.
In all three cases, the slope at the virial radius is $\gamma \approx 2.5$.
It is expected to be less than 3 because of the lower concentration of high-redshift halos relative to the their low-redshift counterparts.
In the runs with star formation, the dark matter has a roughly isothermal inner density profile with $\gamma \approx 2$.
The central dark matter density increases by over an order of magnitude compared with the non-radiative case.

\subsection{Enclosed mass fraction}\label{sec:emf}

\begin{figure*}
	\centering
	\includegraphics[width=\textwidth]{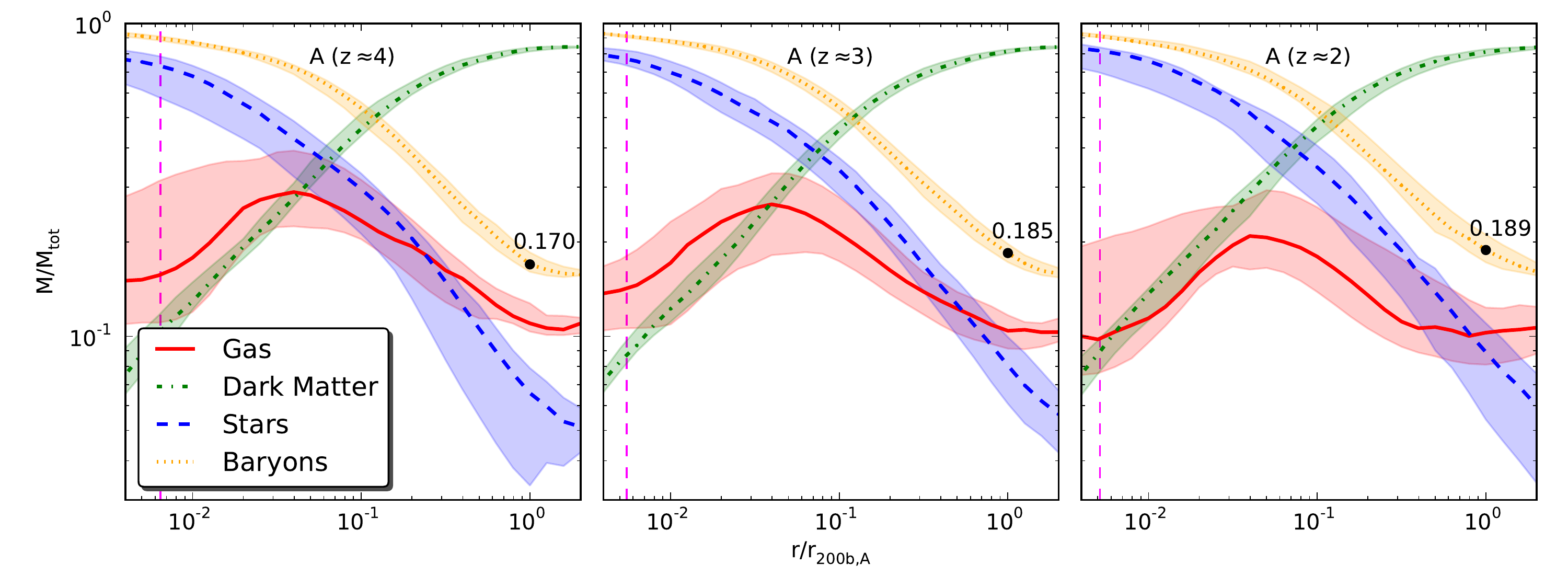}
	\caption{Median enclosed mass fraction $M/M_\mathrm{tot}$ at three epochs in simulation A.
	The median baryon fraction within $r_\mathrm{200b,A}$, $f_\mathrm{B}(r_\mathrm{200b,A})$, is marked with a circle and the value is given in the plot.
	The median resolution scale is marked with a vertical dashed line and the shaded areas mark the region containing 70 percent of the halos.}
	\label{fig:fraction}
\end{figure*}

Figure~\ref{fig:fraction} shows the local enclosed mass fraction $M/M_\mathrm{tot}$ of the different matter species in simulation A.
The fractions sum up to unity at each radius.
Generally, the stars dominate in inner 5\% of the virial radius (about 3 kpc), while
the dark matter dominates in the outer parts of the halo.
The gas mass near the center is typically higher than the dark matter mass.

An interesting feature is the much smaller scatter of the combined baryon profile than of the gas or stellar component separately.
The star formation history varies from object to object, but whenever the stellar density is higher than the median value the gas density is lower by a similar amount, and vice versa.
This is encouraging for our study.
While our modeling of the star formation rate and feedback may not be exactly right, the effect of the baryon dissipation on the dark matter properties is calculated more robustly.

The universal baryon fraction in our simulations is $\Omega_\mathrm{B,0}/\Omega_\mathrm{M,0} = 0.164$.
In general, we find that the median baryon fraction within the virial radius, $f_\mathrm{B}(r_\mathrm{200b,A})$, is higher than universal at all times (by $\approx 5-15\%$) for the runs with cooling and star formation but lower than universal (by $\approx 5\%$) for the non-radiative case.
Also, $f_\mathrm{B}(r_\mathrm{200b,A})$ generally increases with time, up to 15\% above the universal fraction at $z=2$.
The halos in the simulations without supernova feedback can retain slightly more baryons than the halos in the simulations with supernova feedback.
But the effect is relatively weak (between 2\% and 9\%) and within the scatter among individual halos.

These mass fractions should only be used for a qualitative comparison between the different simulations.
There is nothing special about our choice of $r_\mathrm{200b,A}$ as the length scale.
A smooth halo profile extends much further (we plot it out to two virial radii), and in general an edge of a halo is ill defined \citep[see also, e.g.,][]{2006ApJ...645.1001P,2008MNRAS.389..385C}.

\subsection{Concentration}

\begin{figure}
	\centering
	\includegraphics[width=\columnwidth]{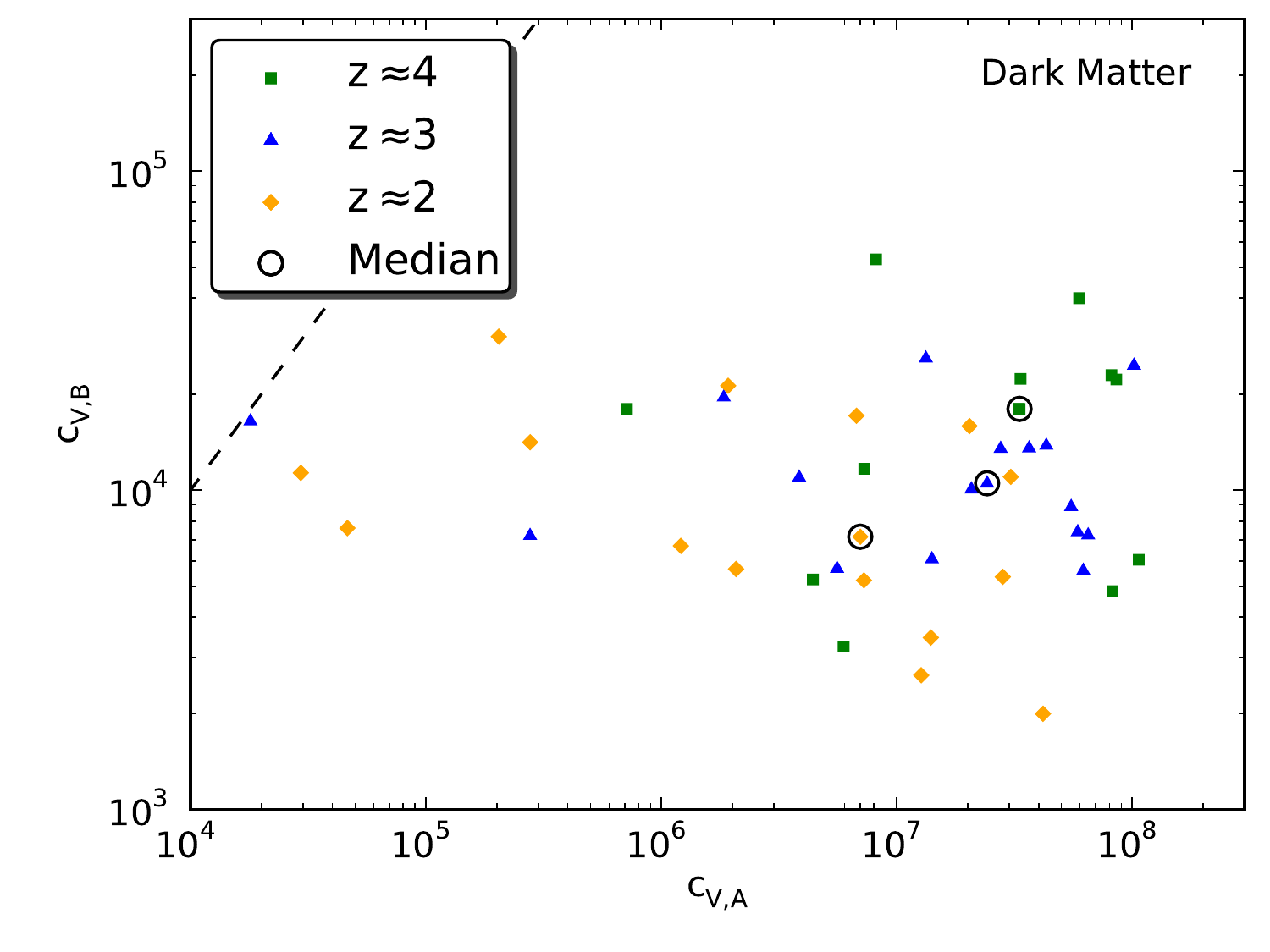}
	\caption{Comparison of halo concentrations $c_\mathrm{V}$ in runs A and B at the three epochs, computed using the dark matter distribution alone.
  Circles show the median concentration of all halos at a given epoch.
	Dashed diagonal line corresponds to the equality relation.}
	\label{fig:concentration}
\end{figure}

In order to describe the more concentrated matter distribution in the simulations with cooling and star formation, we use an intrinsic and general measure for the concentration of a halo.
It is the enclosed density within the radius $r_{\mathrm{max}}$ (location of the peak of the circular velocity curve $V_\mathrm{max}$) in units of the critical density at $z=0$:
\begin{equation}
	c_\mathrm{V} \equiv \frac{\bar{\rho}(r_{\mathrm{max}})}{\rho_{\mathrm{c},0}} = 2 \left(\frac{V_\mathrm{max}}{H_0 ~ r_{\mathrm{max}}}\right)^{2}~.
\end{equation}
This concentration measure has the advantage that it is well defined both for isolated halos and subhalos (as long as the peak of the circular velocity curve can be found) and it does not make any assumptions about a specific shape of the density profile \citep{2002ApJ...572...34A, 2007ApJ...667..859D}.
In principle, $c_\mathrm{V}$ can be derived from observable quantities.
An alternative interpretation is that $c_\mathrm{V}$ is related to the number of rotations, $N_\mathrm{rot}$, at $r_{\mathrm{max}}$ per Hubble time, $1/H_0$, by
\begin{equation}
	N_\mathrm{rot} = \frac{V_\mathrm{max}}{2 \pi r_{\mathrm{max}} H_0} = \sqrt{\frac{c_\mathrm{V}}{8 \pi^2}} \approx 0.113\, \sqrt{c_\mathrm{V}}~.
\end{equation}

Another common measure for a halo concentration is the virial concentration $c_\mathrm{vir} \equiv r_\mathrm{vir}/r_\mathrm{s}$, where $r_\mathrm{s}$ is a characteristic scale radius.
The virial concentration has two main drawbacks: i) $c_\mathrm{vir}$ grows even when the inner mass distribution remains constant, due to the comoving definition of the virial radius, and ii) $c_\mathrm{vir}$ is not well defined for subhalos.
If an analytical halo density profile is known, it is straightforward to calculate the mapping between $c_\mathrm{V}$ and $c_\mathrm{vir}$.
In the case of an NFW profile \citep{1996ApJ...462..563N} see for example Figure 5 in \cite{2007ApJ...667..859D}.
However, in dissipative simulations dark matter no longer follows the NFW profile (the inner parts are modified more than the outer parts) and $c_\mathrm{V}$ is a more useful measure of halo compactness.

Figure~\ref{fig:concentration} shows the concentrations of the objects in run A versus the concentrations in run B.
Here we determine $V_\mathrm{max}$ using the dark matter profile only, because
the total matter distribution is very concentrated and its peak circular velocity is reached below the resolved scale in runs A and A$_\mathrm{NF}$.
Gas cooling and star formation lead to much higher halo concentrations than in the non-radiative case, by a factor of 10 to 10,000.
We also find that the median concentration of the selected halos gradually decreases with time.

\begin{figure}
	\centering
	\includegraphics[width=\columnwidth]{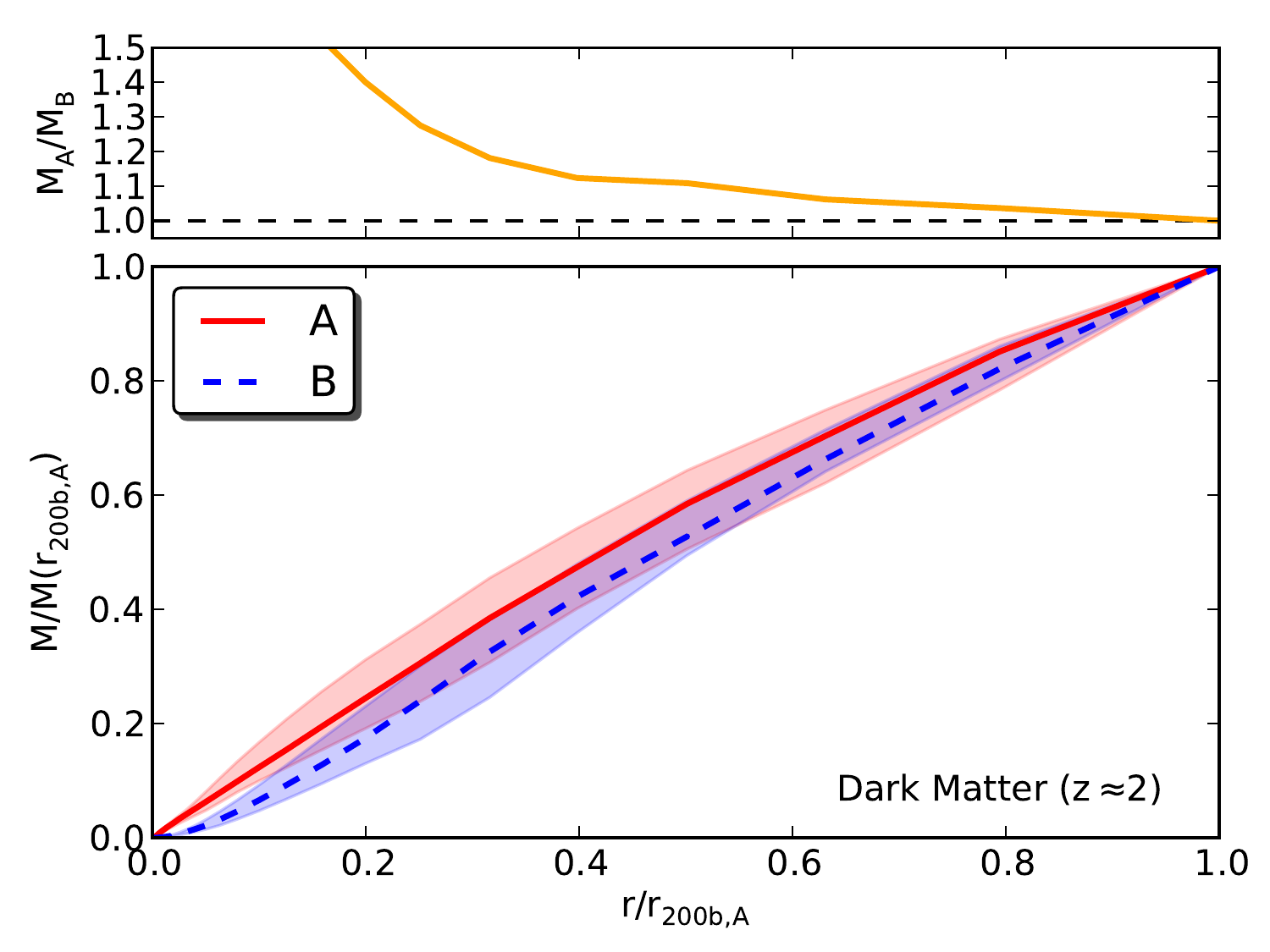}
	\caption{Ratio of the enclosed dark matter mass in the radiative to the non-radiative runs (top panel) and the individual radial profiles $M/M(r_\mathrm{200b,A})$ (bottom panel) at redshift $z \approx 2$.}
	\label{fig:emc}
\end{figure}

While the concentration $c_\mathrm{V}$ is only sensitive to the matter distribution within $r_{\mathrm{max}}$, the processes of gas cooling and star formation affect the dark matter profile throughout the whole halo \citep[see also][]{2008ApJ...672...19R,2010MNRAS.407..435A}.
This is illustrated in Figure~\ref{fig:emc}, where we plot the median enclosed mass fraction $M(r)/M(r_\mathrm{200b,A})$.
The dark matter is enhanced by a factor of around 2 at $\approx 0.1\, r_\mathrm{200b,A}$, and by a still noticeable amount out to a large fraction of the virial radius.

These results indicate that the effects of baryon condensation are not confined to the regions dominated by stellar mass and also lead to subtle (but non-negligible) changes in the mass distribution at larger radii.
A similar effect is also seen in the change of shape, which we discuss in Section~\ref{sec:shape}.

\subsection{Velocity distribution}

\begin{figure}
	\centering
	\includegraphics[width=\columnwidth]{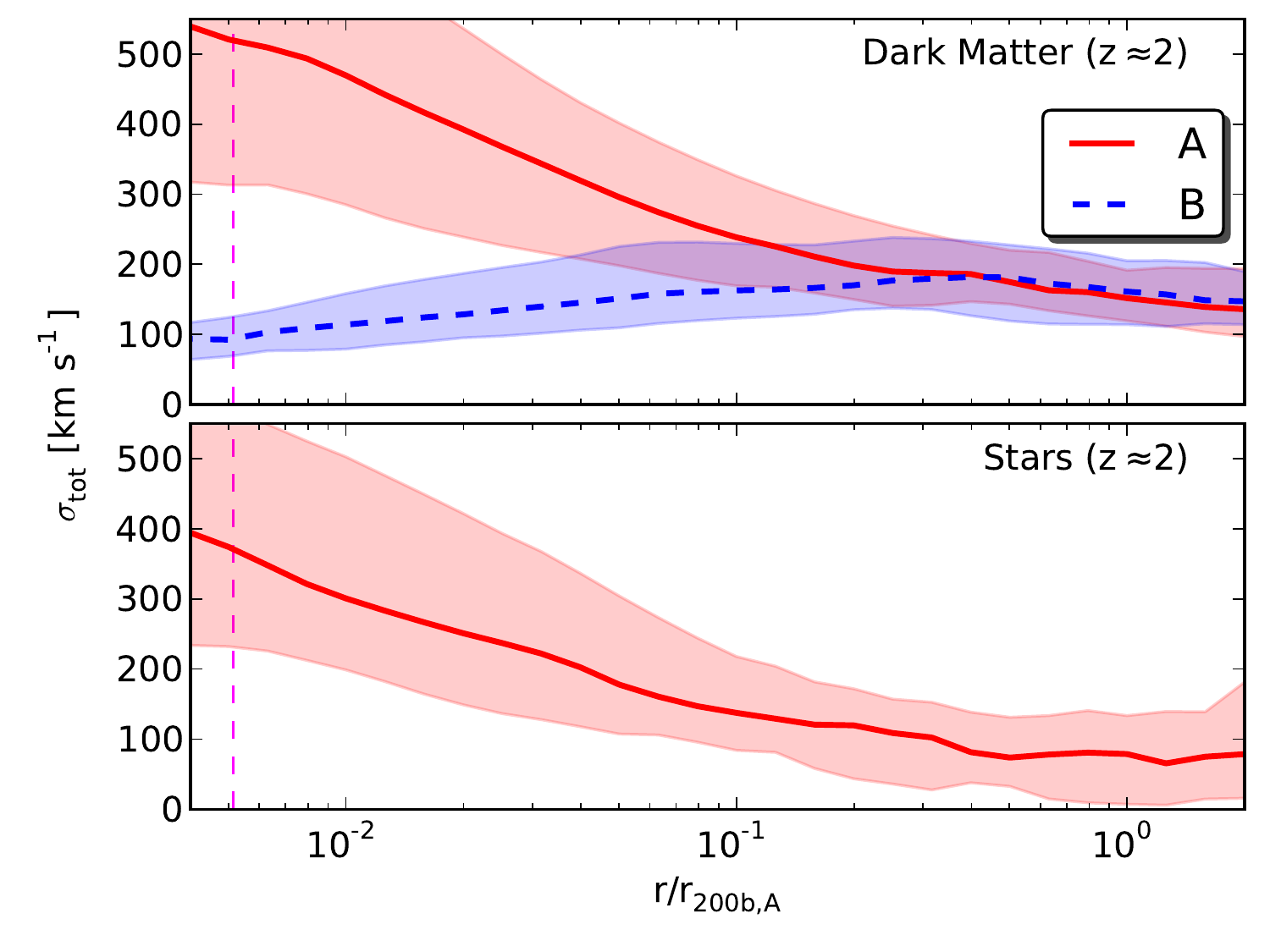}
	\caption{Median total velocity dispersion as a function of radius for the dark matter (top panel) and stars (bottom panel) at $z \approx 2$.
	Stars dominate the mass in the inner $0.07\, r_\mathrm{200b,A}$, but the dark matter dispersion is affected at even larger radii.}
	\label{fig:dispersion}
\end{figure}

In Figure~\ref{fig:dispersion}, we show the total 3D velocity dispersion $\sigma_\mathrm{tot}$ for the dark matter and stars.
In the non-radiative simulation, the dark matter shows a typical velocity dispersion inversion ($\sigma_\mathrm{tot}$ drops towards the center) that is well known from dissipationless simulations.
However, in the runs with cooling and star formation the central velocity dispersion of dark matter increases by a factor $4-5$ and displays a strong negative gradient.
Such a strong increase is due to the concentrated stellar bulge and disk, which dominate the gravitational potential in the inner regions.

\begin{figure}
	\centering
	\includegraphics[width=\columnwidth]{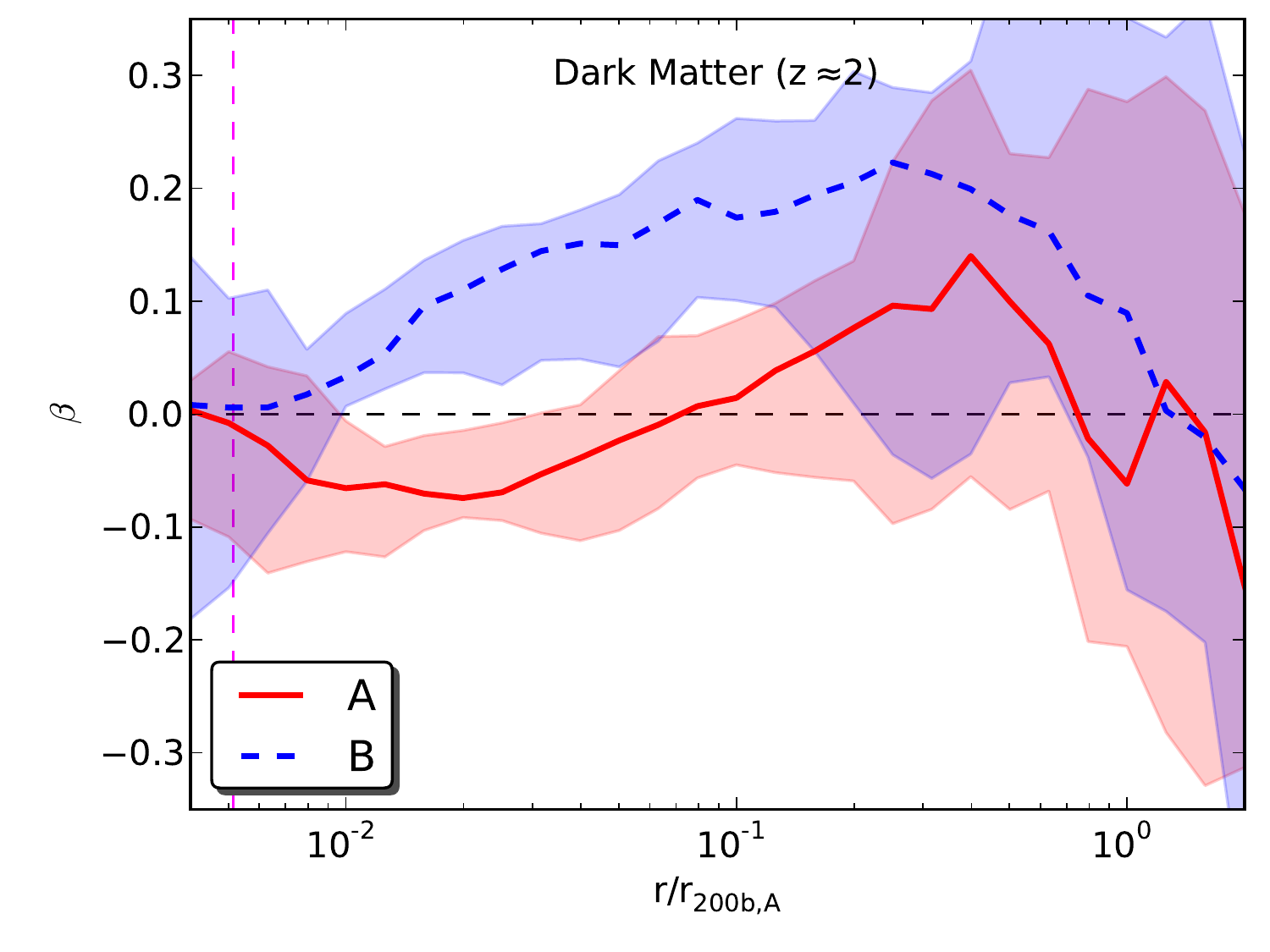}
	\caption{Median anisotropy parameter as a function of radius for the dark matter at $z \approx 2$.}
	\label{fig:anisotropy}
\end{figure}

In order to gain insight into the velocity structure of the halos, we calculate the velocity anisotropy parameter
\begin{equation}
	\beta \equiv 1 - \frac{\sigma_\mathrm{tan}^2}{2 \sigma_\mathrm{rad}^2}
\end{equation}
with $\sigma_\mathrm{rad}$ and $\sigma_\mathrm{tan}$ being the radial and tangential velocity dispersions.
We find that in run B the dark matter has the usual anisotropy profile known from N-body simulations \citep[e.g.][]{2010MNRAS.402...21N}: close to isotropic in the center while becoming more radially anisotropic with radius, with a sharp change towards tangential anisotropy around $r_\mathrm{200b,A}$ (see Figure~\ref{fig:anisotropy}).
In contrast, in our simulation with cooling and star formation the velocity structure is much closer to isotropic over a large radial range.
In the inner regions, around $0.01-0.03\, r_\mathrm{200b,A}$, the dark matter becomes slightly tangentially anisotropic, as it shares the rotation of the stars and gas (and changes its shape, see Section~\ref{sec:shape}).
Only in the outer parts does the velocity dispersion remain radially anisotropic.

\subsection{Pseudo phase-space density}

In dissipationless N-body simulations, the combination of local mass density, $\rho$, and local velocity dispersion, $\sigma$, is shown to follow a power-law over the whole resolved range of radii: $\rho \sigma^{-3} \propto r^{-\alpha}$, with $\alpha = 1.875 - 1.94$ \citep[e.g.][]{1985ApJS...58...39B,2001ApJ...563..483T,2004MNRAS.351..237R,2004MNRAS.352.1109A,2005MNRAS.363.1057D,2007ApJ...671.1108H,2008ApJ...689L..33S,2009MNRAS.398L..21S,2009MNRAS.395.1225V,2010MNRAS.402...21N}.
The quantity $Q \equiv \rho \sigma^{-3}$ is called the ``pseudo phase-space density'' and can be generalized by using different types of the velocity dispersion (total, radial, or tangential) and by treating the power of the velocity dispersion as a free parameter.
\cite{2008ApJ...689L..33S} showed that the value of the slope $\alpha$ depends on these definitions and that, in general, no universal pseudo phase-space density relation exists.
Also, the spherically averaged true phase-space density does not show such a perfect power law behavior \citep{2009MNRAS.398L..21S,2009MNRAS.395.1225V}.

\begin{figure}[t]
	\centering
	\includegraphics[width=\columnwidth]{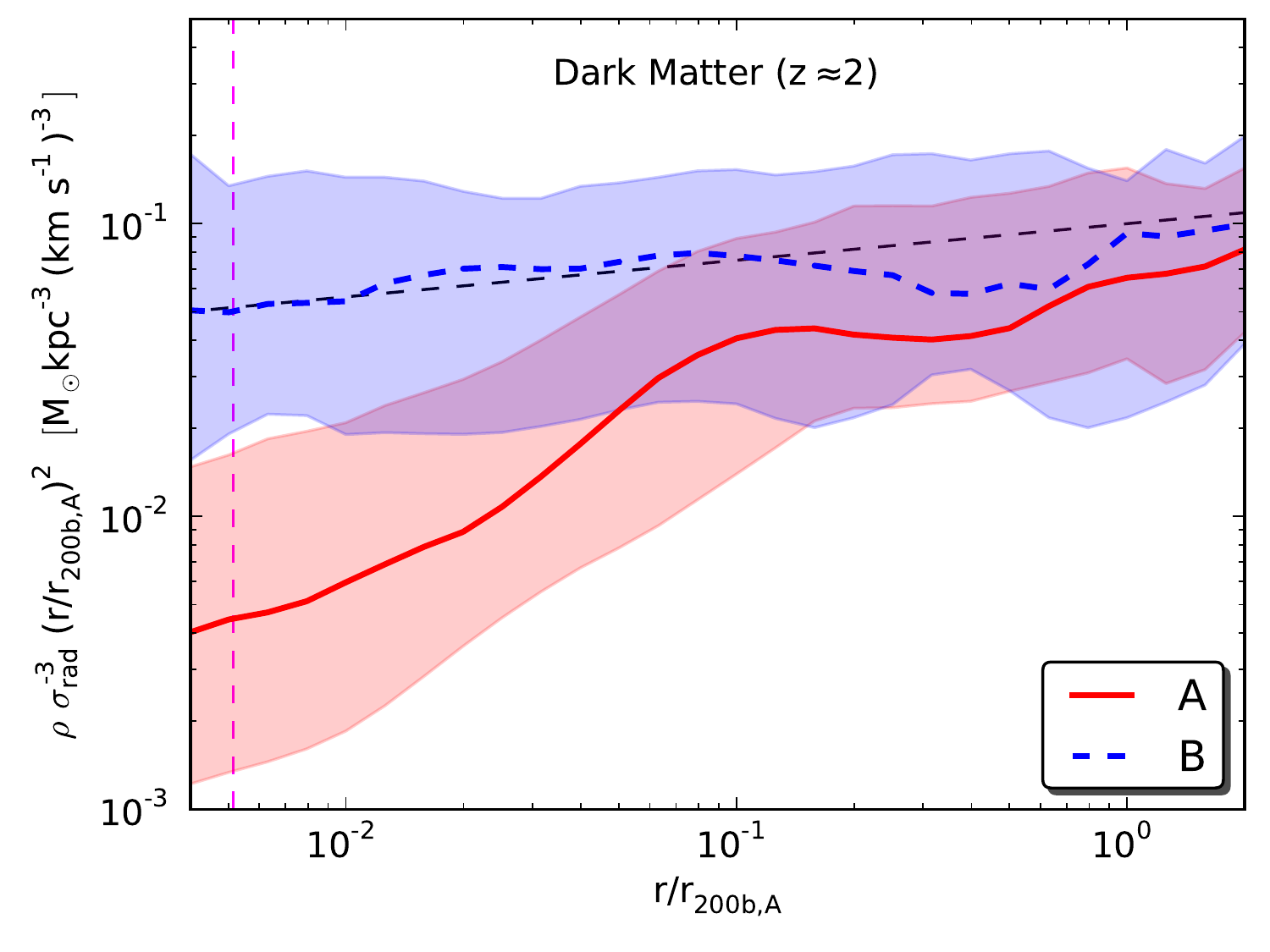}
	\caption{Pseudo phase-space density of dark matter as a function of radius at $z \approx 2$.
	In order to reduce the dynamical range we plot the quantity $\rho \sigma_\mathrm{rad}^{-3} r^2$, similarly to Figure~\ref{fig:density}.
	As a guidance, the dashed line shows the power law with $\alpha = 15/8 = 1.875$, expected from the secondary collapse models \citep{1985ApJS...58...39B}.}
	\label{fig:ppsd}
\end{figure}

Here we investigate how the pseudo phase-space density of dark matter is affected by baryon dissipation.
For illustration, Figure~\ref{fig:ppsd} shows the values of $Q$ defined using the radial velocity dispersion.
In the dissipationless run B we confirm the power law behavior found in previous studies.
The best-fit slope between the resolution radius and $r_\mathrm{200b,A}$ is $\alpha = 1.97$ at $z \approx 2$.
The slope remains essentially constant with time: $\alpha = 1.96$ at $z \approx 3$, and 
$\alpha = 1.91$ at $z \approx 4$, in agreement with the results of \citet{2007ApJ...671.1108H} and \citet{2009MNRAS.395.1225V}.
Qualitatively similar results emerge when using the tangential or total velocity dispersion in the definition of $Q$.
The slope $\alpha$ progressively increases from using $\sigma_\mathrm{tan}$ to $\sigma_\mathrm{tot}$ to $\sigma_\mathrm{rad}$, similar to the findings of \citet{2006JCAP...01..014H} and \citet{2008ApJ...689L..33S}.

However, the pseudo phase-space density is dramatically reduced in the simulation with gas cooling and star formation, by more than a factor of 10 near the center.
The effect is significant all the way to the virial radius, with the sharpest kink at $\approx 0.1\, r_\mathrm{200b,A}$.
A single power law no longer holds in any radial range.
Similar results were obtained also by \citet{2010MNRAS.406..922T}.
Such a strong reduction is due to the increased velocity dispersion of dark matter in the regions dominated by stars and gas, despite the density also increasing (Figure~\ref{fig:density}) but by a smaller amount.
Since the velocity dispersion depends on the details of star formation and feedback, which vary from galaxy to galaxy, our results imply that the pseudo phase-space density profile of dark matter in galaxies is not universal.

\section{Transformation of dark matter halo shape}\label{sec:shape}

In this section, we discuss the \textit{local} shape variation of the mass distribution, i.e. the shape as a function of distance $r$ from the halo center.
We use an iteration method \citep[e.g.][]{1991ApJ...368..325K,1991ApJ...378..496D} and start with a spherical shell with middle radius $r$.
We take logarithmically spaced shells with 7.5 bins per dex in radius.
Then we calculate the shape tensor $\mat{S}$ with the elements
\begin{equation}\label{eq:shapetensor}
  S_{ij} = \frac{\sum_k m_k (\vec{r}_k)_i (\vec{r}_k)_j}{\sum_k m_k}
\end{equation}
where summation is over all particles within that shell, and $(\vec{r}_k)_i$ is the $i$ component of the position vector $\vec{r}_k$ (with respect to the halo center) of a particle/cell $k$ with mass $m_k$.
By diagonalizing $\mat{S}$ we obtain the eigenvectors and eigenvalues at radius $r$.
The eigenvectors give the directions of the semi-principal axes.
The eigenvalues of $\mat{S}$ of an infinitesimally thin ellipsoidal shell are equal to $a^2/3, b^2/3$, and $c^2/3$, where $a$, $b$, and $c$ are the semi-principal axes, with $a \geq b \geq c$.
Hence, the square roots of the eigenvalues are proportional to the lengths of the semi-principal axes.
We then keep the length of the semi-major axis fixed (but the orientation can change) and recalculate $\mat{S}$ by summing over all particles within an ellipsoidal shell with semi-major axis $a=r$ and axis ratios $b/a$ and $c/a$, but with the new orientation.
This iteration is repeated until convergence is reached.
As a convergence criterion we require that the fractional difference between two iteration steps in both axis ratios is smaller than $10^{-3}$.
When referring to the radius or distance from the center for an ellipsoidal shape, we always mean the semi-major axis $a$.

There are many other methods for shape determination used in the literature.
We present tests and comparisons in a companion paper \citep{2011arXiv1107.5582Z}, which further motivates our choice of procedure, since other methods can lead to a significant bias in the measured shape.
For measuring the halo shape it is essential to exclude all particles that are in subhalos.
Not removing subhalo particles can lead to artificially low axis ratios (spikes) at the location of the subhalos (see also \citealt{2011arXiv1107.5582Z}).

\subsection{Axis ratios: from prolate to oblate}

\begin{figure*}
	\centering
	\includegraphics[width=\textwidth]{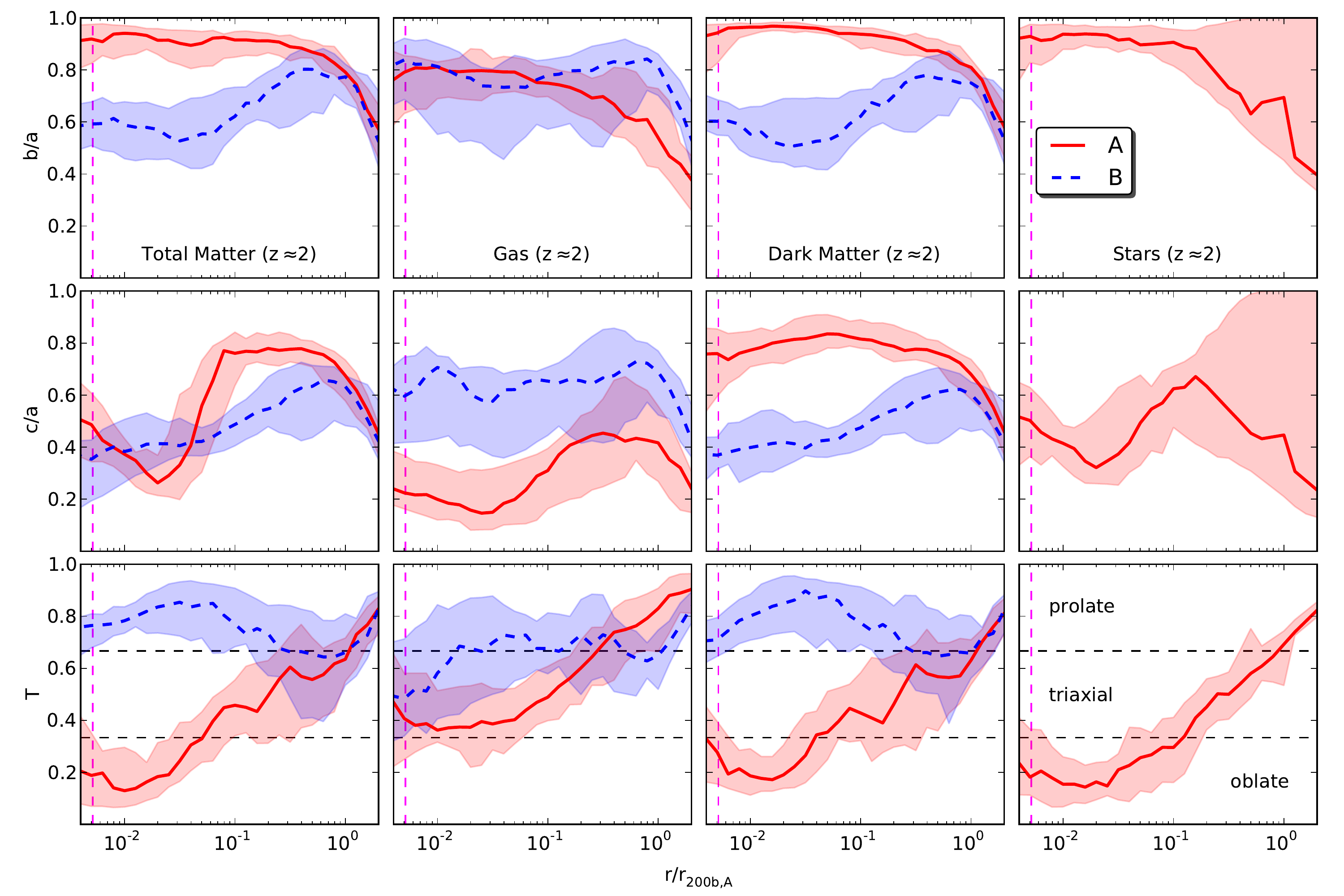}
	\caption{Median axis ratios $b/a$ (top row) and $c/a$ (middle row) as well as the triaxiality parameter $T$ (bottom row) for the different matter components of the halos at $z \approx 2$ as a function of radius.}
	\label{fig:shape}
\end{figure*}

In Figure~\ref{fig:shape}, we show the median axis ratios $b/a$ and $c/a$ as well as the triaxiality parameter
\begin{equation}
	T \equiv \frac{1-(b/a)^2}{1-(c/a)^2}
\end{equation}
for the different matter components at $z \approx 2$.
Ellipsoids are called oblate if $0 \leq T \leq 1/3$, triaxial if $1/3 < T < 2/3$, and prolate if $2/3 \leq T \leq 1$.

In the non-radiative case B, the dark matter shows the well known behavior from N-body simulations where the halo shape is relatively round near the virial radius but becomes progressively prolate (i.e. $c \approx b < a$) towards the center, reaching $b/a \approx 0.55$ and $c/a \approx 0.4$ \citep[see also, e.g.,][]{2005ApJ...627..647B, 2006MNRAS.367.1781A, 2009MNRAS.398L..21S}.

In the dissipative case A, the dark matter shape is much rounder at the center: $b/a \approx 0.95$ and $c/a \approx 0.8$, and the overall shape is oblate instead of prolate.
Such transformation has been seen in previous studies \citep[e.g.,][]{1991ApJ...377..365K, 1994ApJ...422...11E, 1994ApJ...431..617D, 2004ApJ...611L..73K, 2010MNRAS.406..922T, 2011ApJ...734...93L}.
In particular, \cite{2010MNRAS.407..435A} find very similar axis ratios in their smooth particle hydrodynamics (SPH) simulations: $b/a \approx 0.95$, $c/a \approx 0.8-0.9$.
The change in the shape with radius is gradual and is best demonstrated by the triaxiality parameter $T$, which steadily decreases from $T \approx 0.8$ outside the virial radius to $T \approx 0.2$ at the center.

The baryons in run A form a strongly flattened distribution.
The gas settles into a rather thin disk, with $c/a \approx 0.2$ (which includes the hot and warm gas phases; the cold gas disk is still thinner -- see Figure~\ref{fig:disc}), as opposed to an almost prolate elongated shape in run B without dissipation.
The stars also form an oblate disc structure, with a minimum $c/a \approx 0.3$, although it is not as thin as the gaseous disk.
The stellar particles experience gravitational scattering from fluctuation of the potential as soon as they form and therefore diffuse away from the plane of the disk.
Near the center, the stellar distribution becomes less flattened as the disk transitions into a bulge.
The shape of the stellar distribution in the outskirts of the halo has a large scatter because of the low particle number and is not meaningful.

The shape of the total matter distribution is then a consequence of the combination of the individual matter component shapes and their relative importance as a function of radius (shown in Figure~\ref{fig:fraction}).
In the dissipative run A, the shape is approximately round at $r > 0.1\, r_\mathrm{200b,A}$, but sharply turns oblate at the inner radii where baryons dominate the mass \citep[see also][]{2010MNRAS.405.1119K}.
In the innermost region, the matter distribution is determined by the more round bulge.
This detailed shape structure of the mass distribution is important for accurate modeling of the strong gravitational lensing effect due to massive elliptical galaxies, which we discuss in Section~\ref{sec:discussion}.

Over the three epochs that we investigate in detail, we do not find any significant changes in the overall shape of the matter distribution.
Generally, the shape converges farther from center than the density profile.
For example, \cite{2009MNRAS.398L..21S} found that the convergence radius for the shape in a pure N-body simulations was a factor 3 larger than the convergence radius for the density profile.
One should keep this in mind when interpreting the results in Figure~\ref{fig:shape}.

\subsection{Twisting of the halo orientation}

Why does baryon dissipation change the shape of the dark matter distribution so drastically at the center?  
One intuitive interpretation is that dark matter particles respond to the flattened gravitational potential near the disk and transform their orbital structure.
Low-angular momentum box orbits may be replaced by more round tube orbits.
Indeed, we show later in Figure~\ref{fig:angularmomentum_contraction_radius} that the average angular momentum of dark matter particles at the center increases by a factor of several relative to the non-dissipative run.
In addition to studying individual particle orbits, we can conduct a test of this idea based on the global shape of the dark matter halo.
If the oblate spheroid shape of dark matter follows the shape of the baryon disk, the orientation of the inner spheroid should align with that of the disk, regardless of the orientation of dark matter near the virial radius.

We define the $z$-orientation of the baryon disk as the direction of the angular momentum of the gas plus stars in a spherical shell between 1 and 2 kpc (physical) from the center, $\vec{J}_{\mathrm{B},in}$, for each halo individually.
We have chosen to exclude the central 1 kpc region in order to avoid possible complications due to the ill-defined angular momentum of the bulge component.
The exact radial range is not very important and similar ranges around our choice give essentially the same result.

\begin{figure}
	\centering
	\includegraphics[width=\columnwidth]{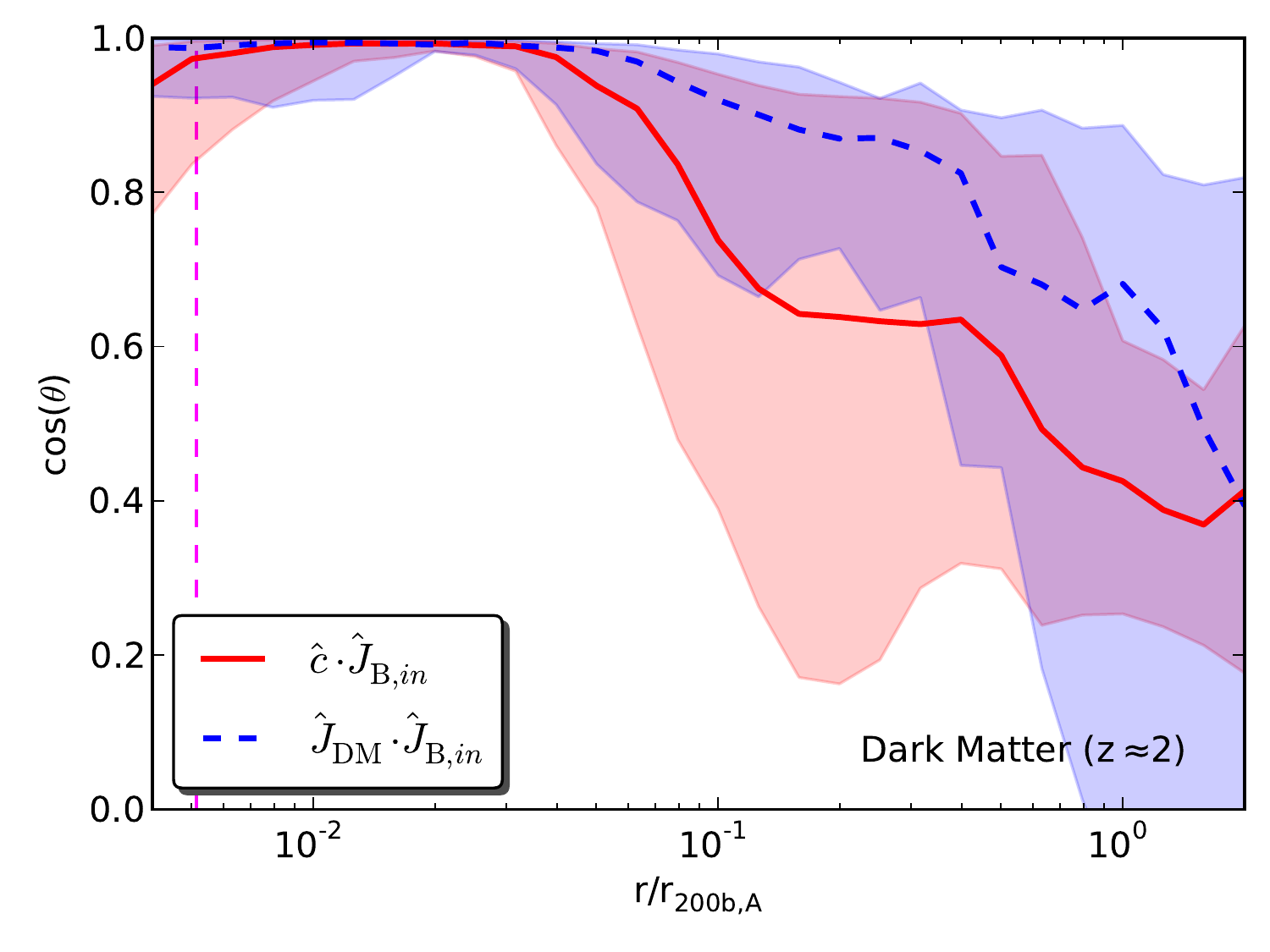}
	\caption{Angle between the direction of the inner angular momentum of the baryons $\hat{\vec{J}}_{\mathrm{B},in}$ (between 1 and 2 kpc) and the direction of the minor axis $\hat{\vec{c}}$ (solid) or the cumulative specific angular momentum $\hat{\vec{J}}_{\mathrm{DM}}$ (dashed) of the dark matter at $z \approx 2$.}
	\label{fig:alignment_SL_z}
\end{figure}

Figure~\ref{fig:alignment_SL_z} shows that the minor axis of the dark matter distribution is almost perfectly aligned with the angular momentum of the disk in the inner 4\% of the virial radius (similar alignment was seen by \citealt{2006ApJ...644..687C}).
However the outer regions of the halo are twisted by $50^\circ$ to $80^\circ$ relative to the disk.
We have confirmed that the orientation of the outer halo is consistent with that in the dissipationless run B.
Therefore, it appears that near the virial radius the dark matter halo retains the shape and orientation set by the large-scale structure formation, while in the inner regions the halo is twisted almost perpendicularly and aligned with the orientation of the baryon disk.

This inner alignment is not just a passive response of dark matter to the disk formation.
\cite{2010MNRAS.408..783R} show that fresh gas feeding the disk is strongly torqued by the hot halo gas on its way into the galaxy center.
Since the hot gas must respond to the shape of the dark matter halo, the mutual alignment is a simultaneous and dynamic process.

The alignment of inner dark matter and stars is also supported by the transformation of the halo velocity anisotropy, from radial to mildly tangential at $r < 0.1\, r_\mathrm{200b,A}$ (see Figure~\ref{fig:anisotropy}).

This flattened, rotating configuration of dark matter near the center resembles what \cite{2009MNRAS.397...44R} named the ``dark disk''.  
In the cosmological simulations of three Milky Way-sized halos with the SPH code GASOLINE, they found a flattened dark matter structure and attributed its origin to the accretion of satellite halos along the directions closely aligned with the stellar disk.
Our simulations similarly include all the cosmological accretion history, with a factor of 5 higher dark matter mass resolution.
We do not see a clear thin disk of dark matter, but rather a smooth oblate distribution near the center, aligned with the stellar disk.
Satellite halos lose most of their mass by tidal stripping at larger radii and do not contribute significantly to the inner regions near the disk.
\cite{2009MNRAS.397...44R} also note the flattening of their halos, with a similar short axis axis ratio $c/a \approx 0.8$.
Thus we agree on the shape and orientation of the inner halo, but our simulations do not reveal any additional ``dark disk''.
The shape of the inner dark matter distribution is determined by the halo contraction, which we discuss in Section~\ref{sec:contra}, and not by accretion of satellites.

Our interpretation is also supported by the results of \cite{2006ApJ...644..687C}, who found that the growth of a galactic bar within an isolated spherical halo led to the transformation of the halo to the oblate shape corotating with the bar.
Their simulations did not include cosmological satellite accretion but nevertheless found the alignment effect similar to our result.

In a wider context, the alignment of the inner dark matter halo is important for the long-term stability of stellar disks.
If the halo remained triaxial, it would scatter stars off the plane of the disk and thus kinematically heat the disk.
An oblate, aligned inner halo is key to the survival of thin stellar disks in CDM cosmology.

A similar picture emerges when studying the orientation of the cumulative specific angular momentum of dark matter, $\vec{J}_{\mathrm{DM}}$.
Near the center the directions of $\vec{J}_{\mathrm{DM}}$ and $\vec{J}_{\mathrm{B},in}$ are well aligned, but at the virial radius they are separated by $47^{+50}_{-20}$ degrees.
Such twisting is in excellent agreement with the AMR simulations of a sample of about 100 galaxies by \cite{2010MNRAS.405..274H}, who found the inner dark matter aligned with the stellar disk to within $18^\circ$, but the outer dark matter misaligned by $\approx 50^\circ$.
The gradual misalignment trend can also be seen in the SPH simulations of \cite{2010MNRAS.404.1137B}, although they only probed it outside $0.25\, r_\mathrm{200b}$ because of lower particle numbers.

These results are also important for evaluating the accuracy of semi-analytical models of galaxy formation that often assume that the specific angular momentum of baryons equals that of dark matter and is conserved during the disk formation.
At the virial radius the angular momentum vectors of the baryons and the dark matter are indeed approximately aligned (within $17^{+30}_{-10}$ degrees), as was found previously in the SPH simulations of \cite{2005ApJ...627L..17B} and in the AMR simulations of \cite{2011arXiv1106.0538K}.
However, the memory of this orientation is completely erased near the disk.
The value of the angular momentum in the inner regions also changes for both baryons and dark matter, as we discuss in Section~\ref{sec:angmom}.

\section{Radial halo contraction}\label{sec:contra}

As we have seen so far, baryon dissipation and condensation near the center lead to a more concentrated dark matter distribution \citep[e.g.][]{1980SvJNP..31..664Z,1984MNRAS.211..753B,1986ApJ...301...27B,1987ApJ...318...15R}.
In this section we describe this enhanced concentration by the modified model of halo contraction \citep{2004ApJ...616...16G,2011GnedinContraction}.

\begin{figure}
	\includegraphics[width=\columnwidth]{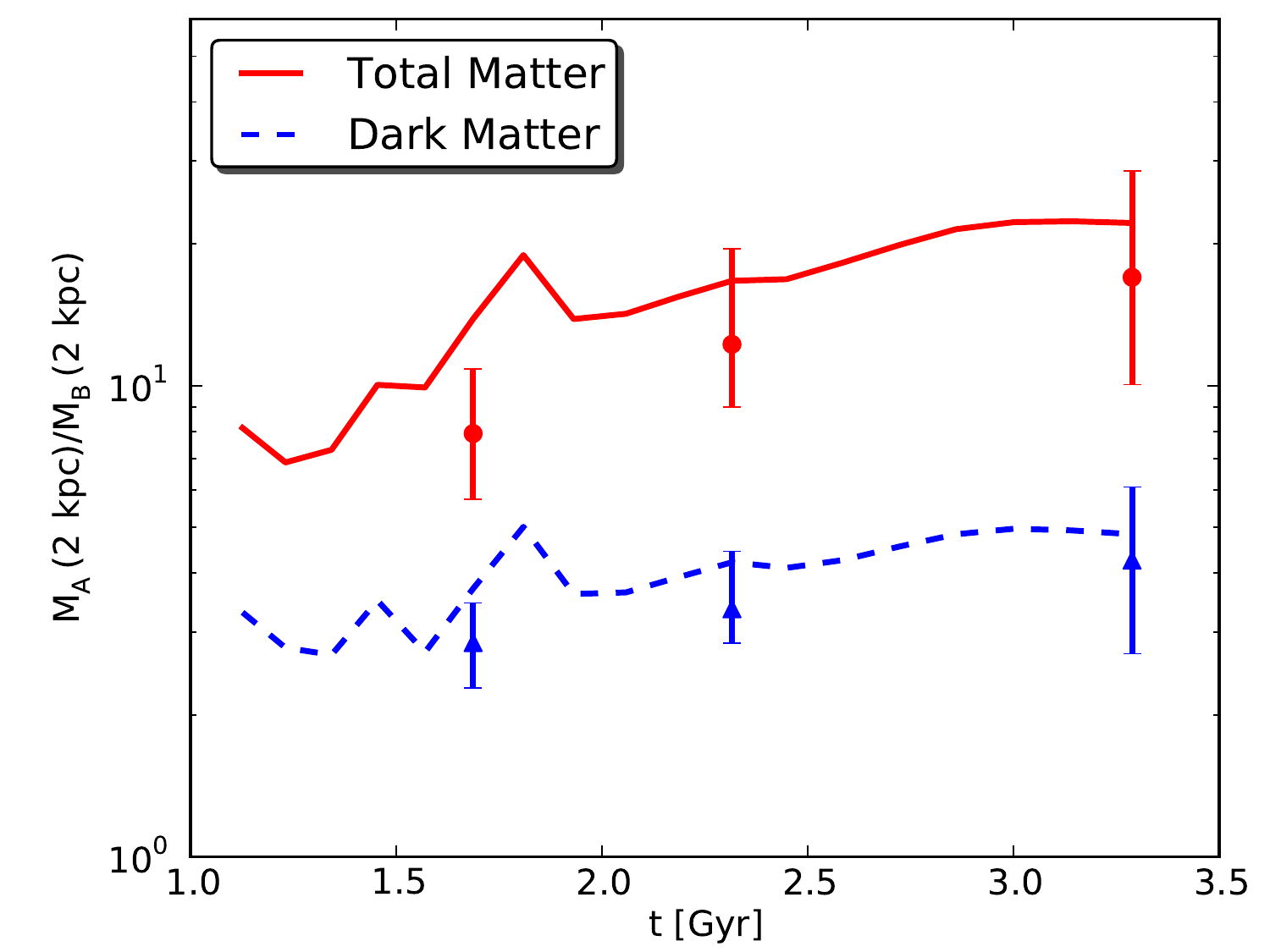}
	\caption{Ratio of the enclosed mass within 2 kpc (physical) in runs A and B.
	Lines are for the main halo.
	For the selected halo sample at $z \approx$ 4, 3, and 2 (corresponding to cosmic times 1.7 Gyr, 2.3 Gyr, and 3.3 Gyr), we plot the median value and the error bars denoting the 15th and 85th percentiles.}
	\label{fig:massratio_time}
\end{figure}

First, we note that the effect of halo contraction is a robust outcome of our simulations, independent of any interpretation using our contraction model.
Figure~\ref{fig:massratio_time} shows the increase of dark matter mass enclosed within a fixed physical radius (2 kpc) compared to the non-radiative case.
The inner dark matter mass is consistently enhanced, by a factor 3 to 6.
Moreover, this enhancement increases steadily with time both for the main halo and in the median of all massive halos.
In addition, dissipation increases the total mass within 2 kpc, baryon plus dark matter, by more than an order of magnitude.

\begin{figure}
	\includegraphics[width=\columnwidth]{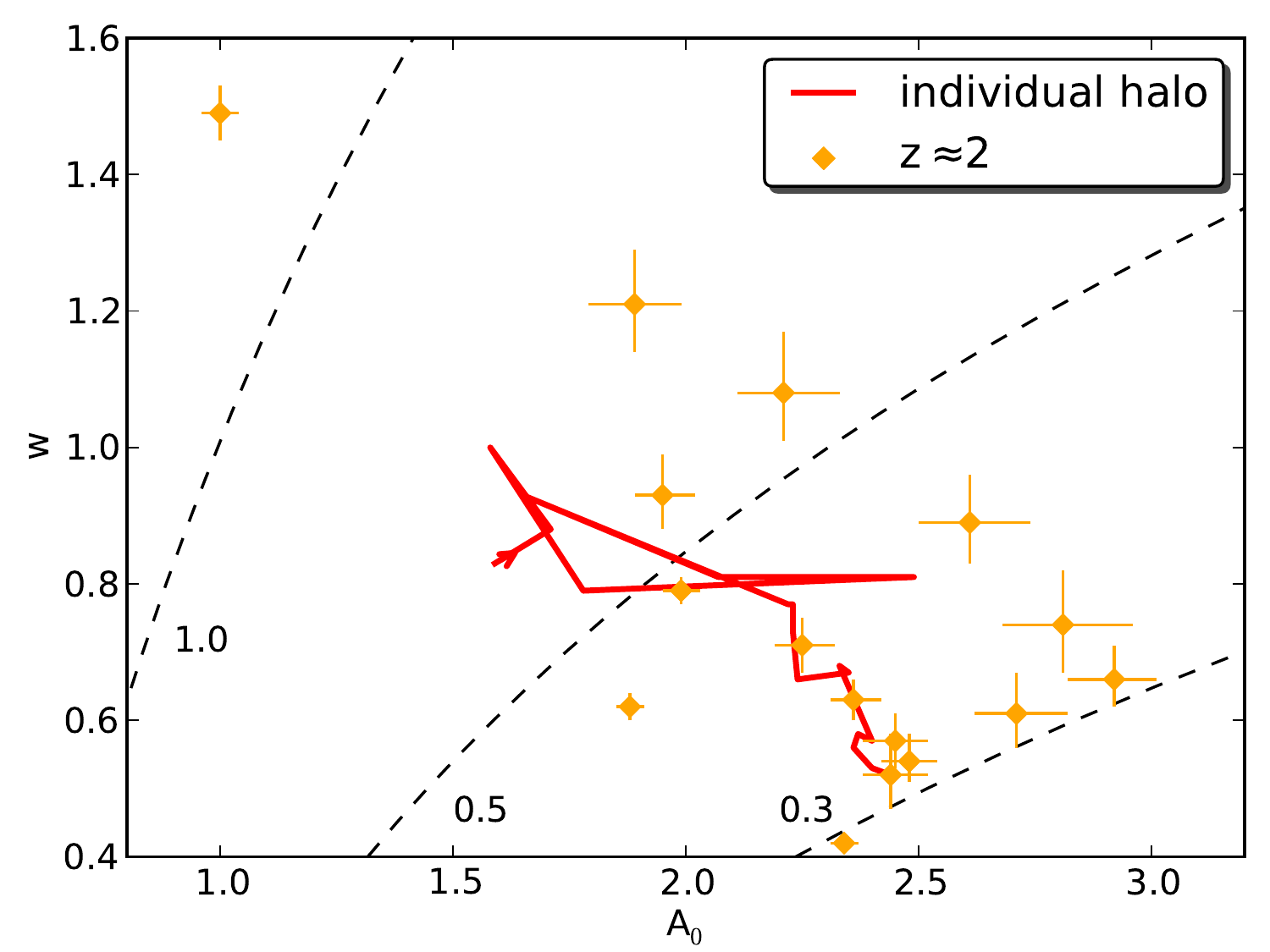}
	\caption{Best-fit values of the parameters of the MAC model for all selected halos at $z \approx 2$.
  Errorbars denote the 90\% confidence limits on each parameter.
  Line shows the track of the main halo from $z\approx 5.2$ to $z\approx 2$.
  The time direction is marked with an arrow.
  Dashed lines show the relation between $A_0$ and $w$ that gives the same mass enhancement factor relative to the SAC model: 100\%, 50\%, and 30\%.}
	\label{fig:contra_A_w}
\end{figure}

Radial halo contraction can be accurately described by the modified adiabatic contraction (MAC) model \citep{2004ApJ...616...16G}, an extension of the standard adiabatic contraction (SAC) model of \citet{1986ApJ...301...27B}.
The SAC model is based on conservation of the specific angular momentum profile of dark matter, $J_{\rm DM}^2(r) = G M(r) r$.
The MAC model is based instead on conservation of the quantity $M(\bar{r}) r$, where $\bar{r}$ is the orbit-averaged radius for particles currently located at radius $r$:
\begin{equation}
  \left[M_{\mathrm{B},i}(\bar{r}_i) + M_{\mathrm{DM},i}(\bar{r}_i)\right] r_i = \left[M_{\mathrm{B},f}(\bar{r}_f) + M_{\mathrm{DM},f}(\bar{r}_f)\right] r_f.
  \label{eq:mac}
\end{equation}
Here $M_{\mathrm{DM},i}$ is the initial dark matter profile, $M_{\mathrm{DM},f}$ is the final dark matter profile (and similar for the baryons), and $r_f$ is the final radius of the dark matter shell that was initially at $r_i$.
The model assumes that contraction of the spherically averaged halo profile can be described as a motion of spherical shells that do not cross each other.
This results in $M_{\mathrm{DM},f}(r_f) = M_{\mathrm{DM},i}(r_i)$.
Using the mass within the orbit-averaged radius $\bar{r}$ approximately accounts for eccentricity of particle orbits in cosmological structure formation simulations.
The orbit-averaged radius can be parametrized as
\begin{equation}
  \frac{\bar{r}}{r_0} = A_0 \left( \frac{r}{r_0} \right)^w~.
  \label{eq:oar}
\end{equation}
This power law relation reflects typical energy and eccentricity distributions of particles in halos.
We take the pivot radius to be $r_0 = 0.03 \, r_\mathrm{200b,A}$, which \cite{2011GnedinContraction} showed to minimize the correlation of $A_0$ and $w$.
The best-fit parameters $A_0$ and $w$ can vary from halo to halo.
The SAC model corresponds to $A_0=1$ and $w=1$.
The amount of contraction typically increases with decreasing values of $A_0$ and increasing values of $w$.

We use a modified version of the code \textit{Contra}\footnote{\url{http://www.astro.lsa.umich.edu/~ognedin/contra/}} that determines the best-fit values for $A_0$ and $w$ by comparing the prediction to the measured dark matter profile in the dissipational simulation.
The fitting procedure is described in \cite{2011GnedinContraction}.

Figure~\ref{fig:contra_A_w} shows the best-fit parameters for our selected halos at $z \approx 2$.
The majority of the halos are clustered around $A_0 \approx 2.3$, $w \approx 0.7$, but some show substantial scatter.
For each individual halo, the MAC model provides an excellent fit of the dark matter profile, with the median fractional mass error of 2.6\% (averaged over all bins).
Overall, the level of contraction is weaker than predicted by the SAC model, confirming the conclusion of \cite{2011GnedinContraction}.

\begin{figure}
	\includegraphics[width=\columnwidth]{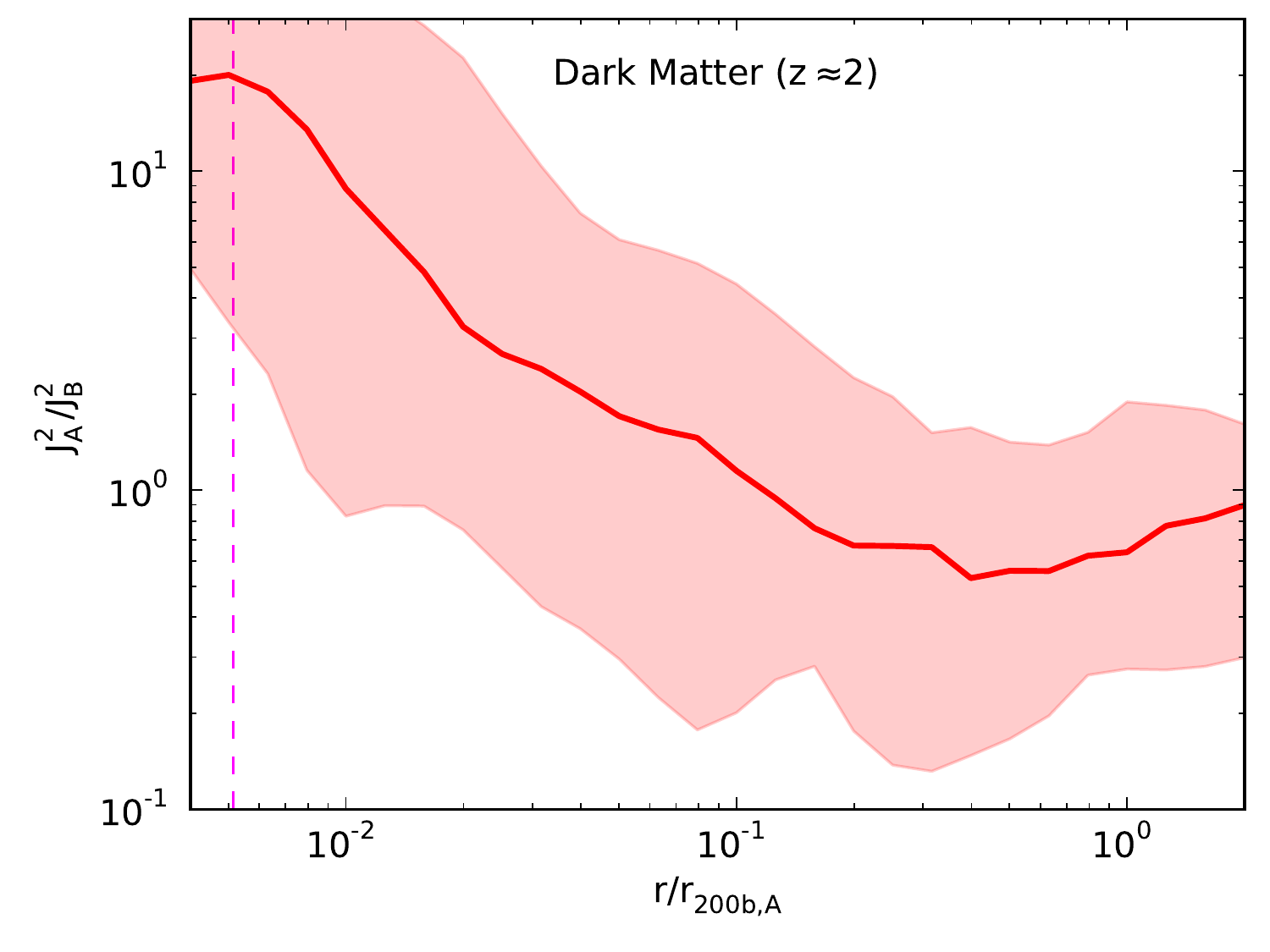}
	\caption{Median ratio of the cumulative specific angular momentum squared of dark matter in runs A and B at $z \approx 2$, for the same amount of matter.
	$J_\mathrm{B}$ is calculated at the radius $r_\mathrm{corr}$ corresponding to the same enclosed dark matter mass as at $r$ in run A, i.e. $M_\mathrm{DM,A}(r) = M_\mathrm{DM,B}(r_\mathrm{corr})$.}
	\label{fig:angularmomentum_contraction_radius}
\end{figure}

Dashed lines in Figure~\ref{fig:contra_A_w} illustrate the strength of halo contraction in comparison to the SAC model.
Let us define the mass enhancement ratio at radius $r$: $F_M(r) \equiv M_{{\rm DM},f}(r) / M_{{\rm DM},i}(r)$.
Each line corresponds to a fixed fraction of this enhancement factor relative to its value in the SAC model: $f_M(r) \equiv F_M(r|A_0,w)/F_M(r|1,1) = 1$, 0.5, and 0.3, respectively.
The radius at which these fractions are evaluated is $r = 0.005\, r_\mathrm{200b,A}$, near our resolution limit.
For most of the halos, the mass enhancement factor at this radius is 30\% to 60\% of the value expected in the SAC model.

Figure~\ref{fig:contra_A_w} also shows the evolution of the main halo over time in the space of the model parameters.
It starts with relatively strong contraction, reaches a peak value around the cosmic time 1.3~Gyr, recedes until 1.8~Gyr, then reaches another peak, and finally settles into a quasi-steady state at $A_0 \approx 2.4$, $w \approx 0.5$.
The two bouts of strongest contraction (when $w$ is highest and $A_0$ is lowest) do not correspond to the epoch of the peak of star formation following a major merger around $t \approx 1.6$~Gyr.
Instead, the first bout precedes the merger and the second happens after the system reaches new dynamical equilibrium.
The exact correlation with the dynamical state is not straightforward or very strong.
At lower redshift the contraction effect stabilizes and is still significant.

\section{Evolution of the angular momentum of dark matter}\label{sec:angmom}

The structure of the halo is affected by the evolution of its angular momentum.
We find that baryon dissipation dramatically affects the angular momentum profile of the inner halo.

At a fixed radius, the total angular momentum $L_{\mathrm{DM}}$ (and the specific angular momentum per unit mass, $J_{\mathrm{DM}}$) increases with time simply because more mass accumulates as the gas falls in and halo contracts.
However, in the idealized scenario of particles on circular orbits and spherical shells that do not cross, the angular momentum of each lagrangian shell with enclosed mass $M_{\mathrm{DM}}$ is conserved, $J^2_{\mathrm{DM}}(M_{\mathrm{DM}}) = G M(r) r = \mathrm{const}$.
As the enclosed baryon mass increases, the shell moves inward but retains the angular momentum profile $J_{\mathrm{DM}}(M_{\mathrm{DM}})$.
This is a test of the foundation of the halo contraction model.
In this section we consider the evolution (in time and between the runs A and B) of the angular momentum of the same spherical shells, that is, the same enclosed amount of material.

Relative to the dissipationless case, the inner dark matter shells gain a lot of angular momentum.
Figure~\ref{fig:angularmomentum_contraction_radius} shows that the median value of $J_{\mathrm{DM}}^2(M_{\mathrm{DM}})$ is greater by as much as a factor of 10 near the center, compared to the angular momentum of the same dark matter mass in run B.
The variance between different halos is large, reflecting their different formation histories.
A similar change of the dark matter angular momentum was also seen in the SPH simulation of \citet{2010MNRAS.404.1137B} at $0.1\, r_{\rm vir}$, the innermost radius they considered.

\begin{figure}[t]
	\includegraphics[width=\columnwidth]{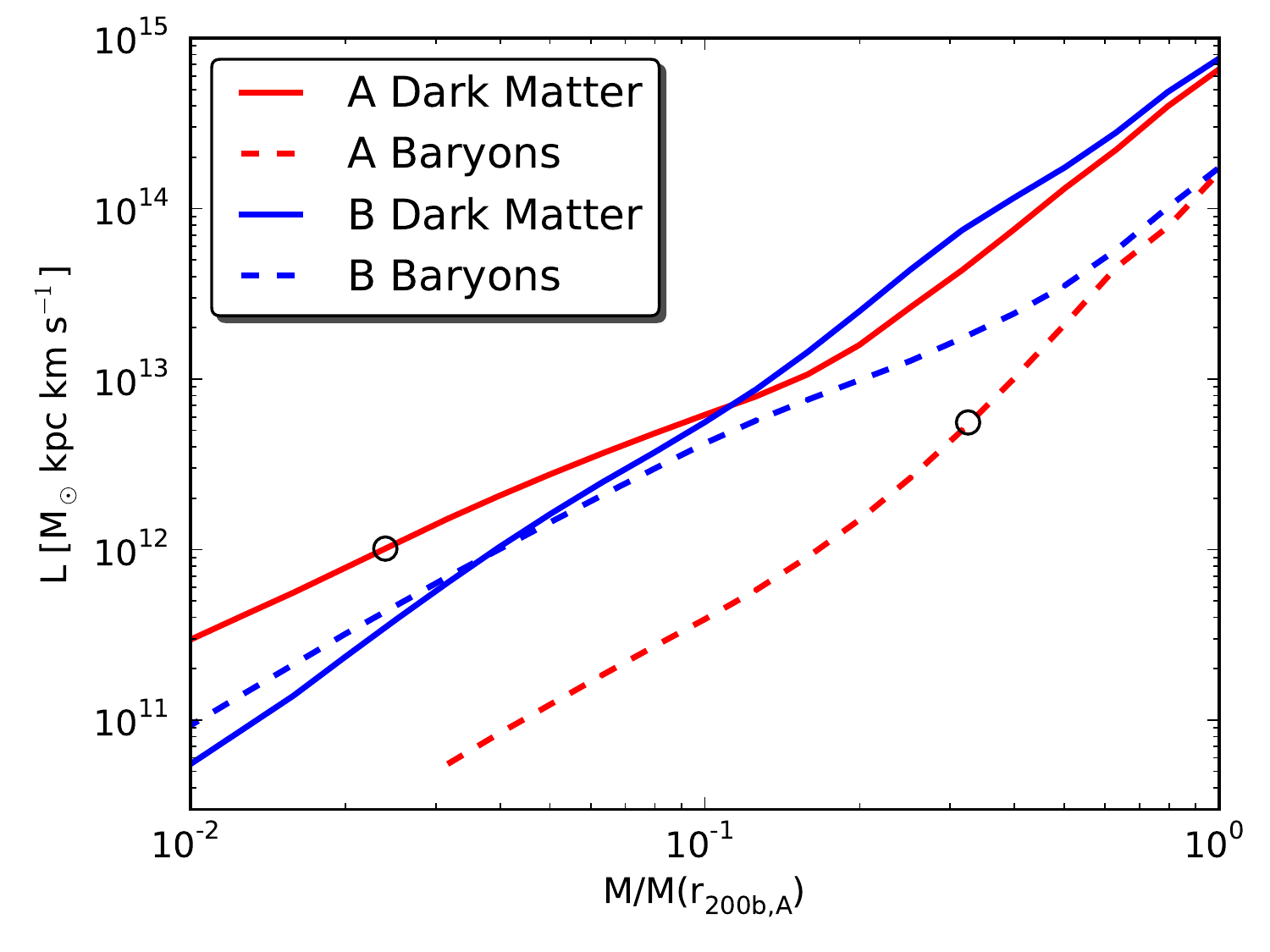}
	\caption{Change of the total angular momentum of dark matter between runs A and B at $z \approx 2$ for the main halo as a function of its enclosed mass fraction rather than radius (\textit{solid lines}).  
  Dashed lines show the same change for the baryons, as a function of the baryon mass fraction.
	One percent mass fraction of dark matter (the left boundary of the plot) corresponds to $r \approx 0.007\, r_\mathrm{200b,A}$, slightly larger than the resolution limit.
  The profile of baryons in run A is plotted to the resolution limit.
  Circles mark the values of the angular momentum within a fixed mass shell in run A at $z \approx 2$ from Figure~\ref{fig:angularmomentum_contraction_time}.}
	\label{fig:angularmomentum_vs_massfraction_cumulative_L_combined}
\end{figure}

\begin{figure}[t]
	\includegraphics[width=\columnwidth]{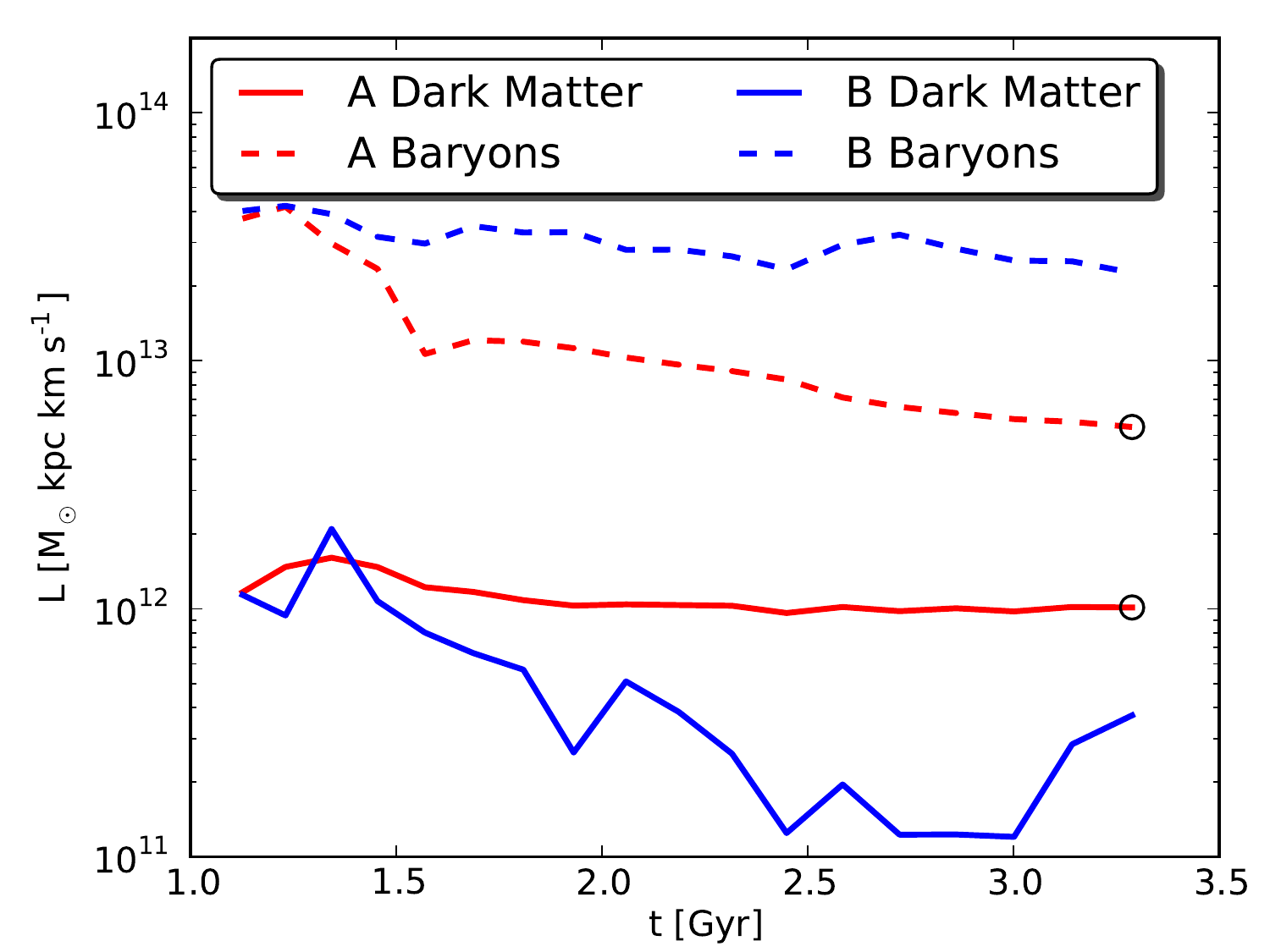}
	\caption{Evolution of the cumulative angular momentum of the dark matter and baryons in the main halo within a radius that contains a constant mass.
	The constant mass for each species was chosen as the enclosed mass within 2 kpc (physical) of the halo at $z \approx 4$ in run A.
  At $z \approx 2$ it corresponds to 32.5\% of the baryon mass and 2.39\% of the dark matter mass (see also marked circles in Figure~\ref{fig:angularmomentum_vs_massfraction_cumulative_L_combined}).}
	\label{fig:angularmomentum_contraction_time}
\end{figure}

To further illustrate this effect, we show in Figure~\ref{fig:angularmomentum_vs_massfraction_cumulative_L_combined} the change of the total angular momentum explicitly as a function of the enclosed mass fraction for both dark matter and baryons.  
In run A the inner 3\% of baryons lose most of the angular momentum ($\approx 90\%$) that they had in run B.
The loss is significant throughout the whole halo, out to the virial radius.
In contrast, the inner 10\% of dark matter gain angular momentum.
Part of this gain may be due to modification of particle orbits near the baryon disk.
Another part may be due to the low-$J_{\mathrm{DM}}$ particles being replaced by the high-$J_{\mathrm{DM}}$ particles from larger radii as a result of mergers and fluctuations of the potential, as first suggested by \cite{2003MNRAS.345..406V}.
Finally, some of the effect may be due to the transfer of angular momentum from the baryons to the dark matter.

However, the comparison between the two runs is masking a strong temporal evolution in the angular momentum profile in each simulation.
A fixed amount of material of both baryons and dark matter, in both runs, systematically loses the angular momentum in the inner galaxy.
Figure~\ref{fig:angularmomentum_contraction_time} shows this evolution in the main halo for the shell containing 2.39\% of the dark matter mass and 32.5\% of the baryon mass at $z\approx 2$ (it has radius of $\approx$ 1.5 kpc (dark matter) and $\approx$ 0.5 kpc (baryons), respectively, at that epoch).
The sharpest drop of $L_{\mathrm{B}}$, by a factor of 2, is around the cosmic time of 1.6 Gyr.
This epoch corresponds to a major merger event, when even the mass within the inner several kpc increases noticeably.
It causes also the largest burst of star formation in this galaxy, with the rate exceeding $100 \, \Mo$~yr$^{-1}$.
It has been known since the work of \cite{2002ApJ...581..799V} that the spin of the halos decreases after mergers.
A non-negligible fraction of particles near the center gains enough energy to move outside the virial radius \citep[e.g.,][]{2006ApJ...641..647K}.
Later, \cite{2007MNRAS.380L..58D} showed that out-of-equilibrium halos have, on average, higher spin than relaxed systems, suggesting that the post-merger virialization process leads to a net decrease in the halo spin.
Our results show that this effect may operate even at radii as small as a few kpc, where the high-$J_{\mathrm{B}}$ gas moves to larger radii and leaves the remaining baryons with much lower angular momentum.

Figure~\ref{fig:angularmomentum_contraction_time} shows that without baryon dissipation, the enclosed $L_{\mathrm{DM}}(M_{\mathrm{DM}})$ of a fixed dark matter mass $M_{\mathrm{DM}}$ also decreases with time.
In contrast, as a result of gas cooling and star formation in run A, the angular momentum of the same mass, $L_{\mathrm{DM}}(M_{\mathrm{DM}})$, remains relatively constant over time.
This inner region contains a much larger fraction of baryon mass than dark matter mass, and therefore baryons carry much higher total enclosed angular momentum, $L_{\mathrm{B}} \gg L_{\mathrm{DM}}$.
The lost baryon angular momentum is more than enough to feed the angular momentum of the inner dark matter.
Note that this result implies that at a fixed physical radius $L_{\mathrm{DM}}(r)$ increases over time, because the enclosed mass $M_{\mathrm{DM}}(r)$ increases as the halo contracts.

\begin{figure}[t]
	\includegraphics[width=\columnwidth]{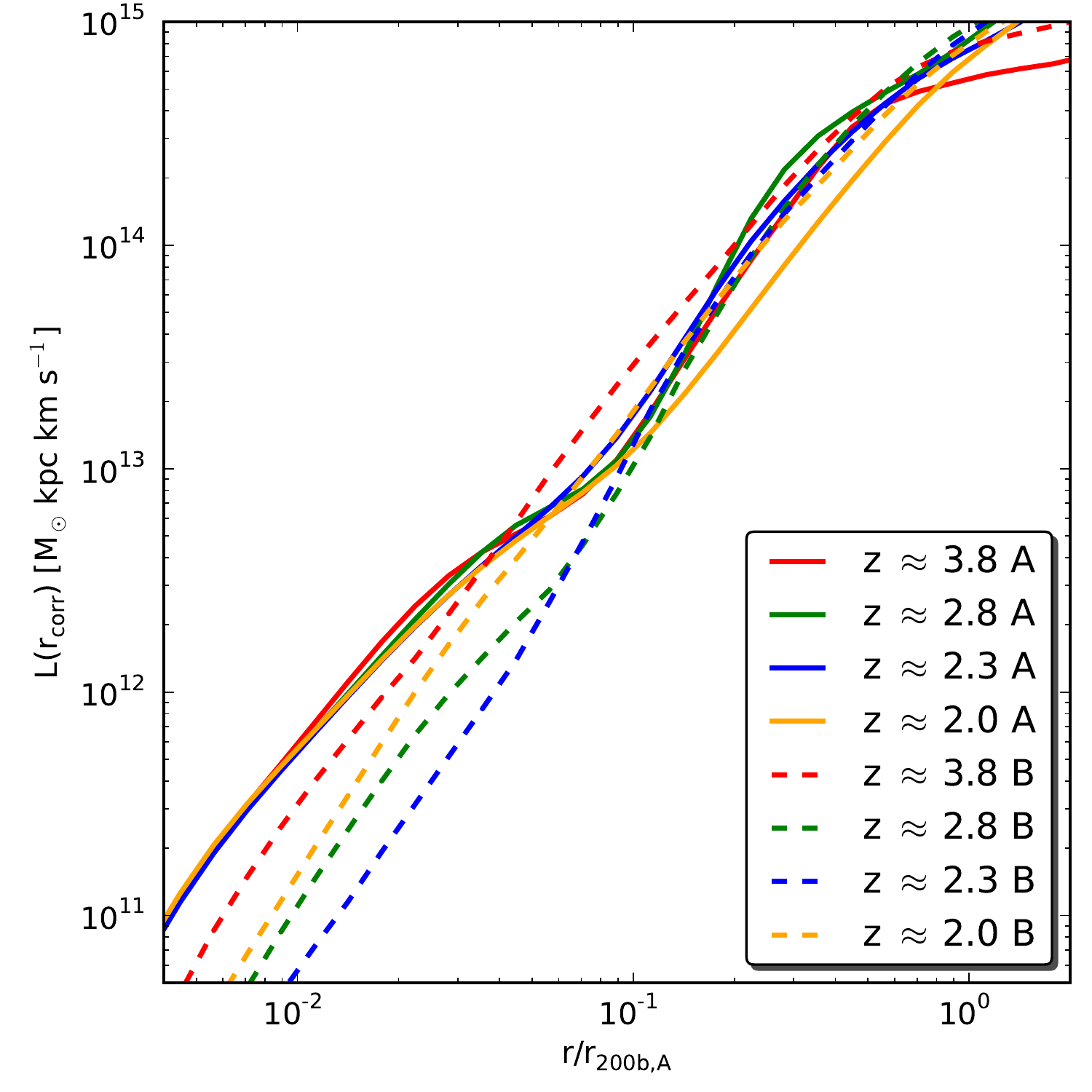}
	\caption{Angular momentum profiles of the dark matter shells in the main halo.
  For each radius $r$ at $z \approx 2$ we matched the same enclosed mass at previous epochs.
  Solid lines are for run A, dashed lines are for run B.}
	\label{fig:angularmomentum_time_series}
\end{figure}

The remarkable constancy of the angular momentum in run A extends to all dark matter shells out to $\approx 0.1\, r_\mathrm{200b,A}$.
Figure~\ref{fig:angularmomentum_time_series} shows that the inner angular momentum profile has decreased by less than $10-20\%$ between $z\approx 4$ and $z\approx 2$.
At higher $z > 4$, when the central baryon accumulation was dynamically insignificant, the inner $L_{\mathrm{DM}}(M_{\mathrm{DM}})$ was fluctuating by a larger amount.
Since $z\approx 4$, the dense stellar bulge and disk have stabilized the gravitational potential and helped maintain a very constant angular momentum profile of the dark matter shells.
In contrast, the inner $L_{\mathrm{DM}}(M_{\mathrm{DM}})$ of dark matter in the non-radiative run B decreased overall by a factor of several.

Did baryons transfer just enough angular momentum to inner dark matter to balance the loss due to mergers?
Details of the angular momentum evolution still need to be investigated in future studies, but the overall effect illustrated in Figure~\ref{fig:angularmomentum_time_series} is new and significant.
The angular momentum profile of inner dark matter shells is fortuitously constant in the region dominated by the baryons, at least to the final epoch of our simulation.
This effect lends strong support to the underlying equation in the models of halo contraction.
In addition, the radial mixing and eccentricity of particle orbits strengthen the motivation for using the MAC model, which is based on even more robust conservation of the radial action.

\section{Discussion}\label{sec:discussion}

\subsection{Comparison with observations}

At the moment there are only a few observational constraints at high redshift.
\cite{2009Natur.460..717V} report on a massive compact galaxy at redshift $z = 2.2$ with the stellar mass around $2 \times 10^{11}\, \Mo$ and the line-of-sight velocity dispersion of $510^{+165}_{-95}$ km s$^{-1}$.
Converting this one-dimensional dispersion to the three-dimensional dispersion with the factor $\sqrt{3}$ gives roughly 900 km s$^{-1}$.
Our most massive halo in run A at $z \approx 2$ has total stellar mass of $1.2 \times 10^{11}\, \Mo$ and a stellar 3D velocity dispersion around 725 km s$^{-1}$ (at $0.005\, r_\mathrm{200b,A}$).
It is reasonably consistent with the observed analogue.

\cite{2009ApJ...703L..51K} determine the inner slope of the total matter density for a sample of galaxies from the Sloan Lens ACS (SLACS) survey.
They report an essentially constant slope $\gamma = 2.09^{+0.03}_{-0.02}$ in the inner region ($0.2-1.3\, R_e$) in the redshift range $z \approx 0-0.4$.
The median slope for our galaxies is $\gamma \approx 2.5$ in the same region (Figure~\ref{fig:slope}) and it does not change substantially between $z \approx 4$ and $z \approx 2$.
In this inner region most of the gas has already been converted into stars, which dominate the central mass (Figure~\ref{fig:fraction}).
If the inner density slope does not decrease at $z<2$, as may be expected for collisionless systems which preserve the steepest density profile during a merger \citep[e.g.,][]{2005MNRAS.360..892D, 2006ApJ...641..647K, 2008MNRAS.386.1543Z}, then our simulated galaxy profiles would be too steep.
However, continuous star formation and accretion of satellites may decrease the central concentration and build up the extent of the stellar disk, thus reducing the overall density slope.
The mass profile of the lensing spiral galaxies will soon be available from the
new multi-band SWELLS survey \citep{2011arXiv1104.5663T}.

If we fit an exponential profile to each stellar disk of our selected objects at $z \approx 3$ in the range between 1 and 3 kpc, we obtain a median exponential scale length $R_d = 470^{+120}_{-65}$~pc.
Since the stellar profiles are not strictly exponential, a more robust measure of the stellar distribution is an effective half-mass effective radius, which we find to be $R_e = 800^{+200}_{-110}$~pc.
This size in reasonable agreement (and even larger) with \cite{2011ApJ...736...48R}, who found an effective radius $R_e \approx 600$~pc for their stacked sample of Lyman break galaxies at $z \approx 3$ in the same radial range.
Our size also agrees with the Keck/OSIRIS integral field spectroscopy study of 12 star-forming galaxies at $z \approx 2-3$ by \cite{2009ApJ...697.2057L}, who found the radii of ionized nebula emission between 0.6 and 1.6 kpc.
However, \cite{2009ApJ...706.1364F} found substantially larger effective radii of the H$\alpha$ emitting region (between 1 and 8 kpc) in their study of 80 star-forming galaxies at $z \approx 1-3.5$ using VLT/SINFONI integral field spectroscopy.
We will discuss the detailed structure of the disks in our galaxies in a future publication.

\cite{2011arXiv1106.2816M} showed that the half-light size of Lyman-$\alpha$ emitting galaxies does not change between redshift $z \approx 6$ and $z \approx 2$ ($R_e \approx 1$ kpc) in contrast with the sizes of the Lyman-break galaxies which are increasing over the same period by a factor of 3.  
While both types of galaxy are actively star forming, the Lyman-$\alpha$-selected objects are typically younger, relatively less massive, and less evolved chemically.  
The sizes of our simulated galaxies are consistent with the Lyman-$\alpha$ emitters.  
Since Lyman-breaks are picking moderate-aged stars, it is an indication that the size of stellar disks/bulges systematically increases with cosmic time.

\subsection{Comparison with simulations}

The problem of a very concentrated stellar distribution is prominent in most simulations of galaxy formation \citep[e.g.][]{2007MNRAS.374.1479G,2009MNRAS.396..696S,2010MNRAS.408..812S}, although two recent SPH simulations seem to have overcome this by using very strong feedback with a supernova blastwave and applying a high star formation density threshold, for a dwarf galaxy \citep{2010Natur.463..203G} and a large galaxy \citep{2011arXiv1103.6030G}.  For example, \citet{2010Natur.463..203G} had gas cooling shut off in a region of $0.1-0.3$ kpc in radius for $5-10$ Myr after the supernova explosion, where most of the energy is lost radiatively in our simulations.

\cite{2009ApJ...692L...1J} performed cosmological simulations of galaxies at $z=3$ with the AMR code Enzo.
Their simulated galaxies are in a similar mass range to our sample and also show very centrally concentrated stellar profiles, despite substantially different treatment of gas physics and star formation.
We have also checked that in our simulations the dense stellar core was already built up at earlier epochs, at least at $z=5$.
\cite{2009ApJ...692L...1J} suggested possible ways around this problem: an early reionization to suppress gas condensation; a very strong feedback from stars or central black holes to reduce overall star formation; or a small-scale cutoff in the primordial fluctuation power spectrum.

However, \cite{2011MNRAS.410.1391A} was able to form a more extended disk galaxy at $z=0$ using the AMR code RAMSES.
They performed a parameter study with the star formation efficiency per free-fall time varying between $\epsilon_\mathrm{ff} = 0.01$ and 0.05, and found that decreasing $\epsilon_\mathrm{ff}$ led to a less concentrated stellar distribution, despite smaller stellar mass and weaker feedback.
We use an even smaller star formation efficiency ($\epsilon_\mathrm{ff} = 0.007$) but find steeper stellar profiles.
This discrepancy is probably due to a combination of two effects.
First, the \cite{2011MNRAS.410.1391A} simulation has lower spatial resolution (340 pc vs. 90 pc in our run A at $z=2$) and forms stars based on the \textit{total} gas density rather the \textit{molecular hydrogen} density, and at much lower density than in our case.
Molecular hydrogen naturally forms only at high gas density and therefore in our simulations star formation takes place in systematically more concentrated regions (see also Figure~\ref{fig:disc}).
For any given feedback model, the feedback is less efficient in our simulations.
Second, \cite{2011MNRAS.410.1391A} analyze the stellar disk at $z=0$, as opposed to $z \approx 2$ in our case.
The high-redshift disks are expected to grow gradually in size and may become less concentrated by the present epoch.

\section{Summary}

We have presented a study of the impact of baryon physics on the matter distribution in galaxy formation simulations.
An important new feature of our simulations is the modeling of the physics of molecular hydrogen and the star formation prescription based on the local \textit{molecular hydrogen} density and not on the \textit{total} gas density.
We investigated the properties of the most massive halos with $M_\mathrm{200b} \geq 10^{11}\, \Mo$ to redshift $z \approx 2$, using three simulations of the same cosmological volume that differ by the included gas physics and stellar feedback.

The main conclusions of this work are:

\begin{itemize}
\item The spatial distribution of the molecular gas, and the stars that form from it, is very compact.
The median half-mass size of the stellar disks is 0.8 kpc at $z \approx 3$.
The molecular disk is much thinner than the distribution of the neutral atomic gas.

\item Compared to the matching dissipationless run, the inner regions of the dark matter halo change shape from prolate to oblate and align with the orientation of the stellar disk.  
This alignment of the inner halo is important for the long-term stability of the thin stellar disk.

\item Despite the alignment, the inner halo is only mildly oblate: the short-to-long axis ratio is $c/a \approx 0.8$.  
We do not detect a ``dark disk'' of comparable thinness to the stellar disk.

\item The radial profile of the dark matter halo contracts in response to baryon dissipation.  
In the inner 10\% of the virial radius, the logarithmic slope of the dark matter density profile is constant, $\gamma \approx 2$.
Halo mass within the inner 2 kpc is consistently enhanced in all galaxies, by a factor of 3 to 6.  
The mass enhancement increases steadily with time.
The modification of the halo profile is accurately described by the modified model of halo contraction.

\item Baryon dissipation affects the structure of the dark matter halo far outside the extent of the stellar disk.
The enclosed mass is enhanced by $\approx 40\%$ at $0.2~r_\mathrm{200b}$.
The halo concentration is also significantly increased relative to the non-radiative case.
The halo velocity distribution is consistent with being isotropic throughout the whole halo.

\item The pseudo phase-space density profile of dark matter is not universal and cannot be described by a single power law.
In the inner region dominated by stars, the phase-space density is reduced by an order of magnitude relative to the dissipationless case.

\item In the non-radiative run, a fixed amount of inner dark matter loses most of its angular momentum between $z=5$ and $z=2$.
This angular momentum is transferred outward during major mergers and potential fluctuations.
In contrast, in the galaxy formation run the specific angular momentum of dark matter is approximately constant in time, while the baryons lose most of theirs.
Some of the baryon angular momentum is transferred to the inner dark matter, helping to offset the loss during major mergers.
The approximate conservation of the dark matter angular momentum may explain why the halo contraction model describes the modification of the halo profile so well, despite the simplicity of the model and complexity of the galaxy formation process.
\end{itemize}

\acknowledgements
\lastpagefootnotes
MZ, OYG, NYG, and AVK are supported in part by NSF grant AST-0708087.
The simulations and analysis in this work have been performed on the Joint Fermilab-KICP Supercomputing Cluster (supported by grants from Fermilab, Kavli Institute for Cosmological Physics, and the University of Chicago) and the Legato and Flux clusters at the University of Michigan.
This research has made use of NASA's Astrophysics Data System (ADS), the arXiv.org preprint server, the visualization tool VisIt, the computer algebra system Maxima, and the Python plotting library Matplotlib.

\appendix
\section{Profiling routine}

In order to determine the properties of simulated galaxies, we have built a profiling routine that takes the phase-space coordinates from any halo finder as input.
If necessary, the routine can recenter the phase-space coordinates of a galaxy through a shrinking sphere method \citep{2003MNRAS.338...14P}.
The profiling routine then creates a radial grid between the specified boundaries $r_\mathrm{gridmin}$ and $r_\mathrm{gridmax}$.
The inner boundary is set by the resolution scale of the simulation, and we chose a value of $r_\mathrm{gridmin} = L_\mathrm{box}/(256 \cdot 500) = 512/500\,  L_9$ = 286 pc (comoving).
The outer boundary is set individually for each halo based on an estimate of the virial radius $r_{\mathrm{200b},i}$ of halo $i$.
Here we set $r_{\mathrm{gridmax},i} = 5 \, r_{\mathrm{200b},i}$.
For the binning, we specify a number of bins per dex in radius.
This has the advantage that it is adaptive to the size of the individual objects: larger objects will have more bins than smaller objects.
It is also more memory efficient than using the same number of bins for all objects.
Additionally, one can choose appropriate binning depending on the resolution of the simulation.
For the analysis presented in this paper we chose 15 bins per dex in radius.

In order to reduce the memory load of the profiling routine, we read only parts of data at once: generally $10^7$ particles or cells.
This allows us to analyze even biggest simulations on a desktop computer with a reasonable amount of memory.
Also, the halo catalog can be split up, although it is not needed in most cases.
In order to speed up the assignment of particles into the bins, we use a chaining mesh method and linked list data structures \citep[e.g.][]{1988csup.book.....H}.
By choosing the chaining mesh size of the order of the (largest) halo size in the simulation, we only need to check particles in the neighboring mesh cells of a halo for being within $r_\mathrm{gridmax}$ of that halo.

The profiling routine calculates radial profiles for each individual matter species as well as for the total matter distribution.
By searching for a maximum in the circular velocity curve on the grid we determine $r_\mathrm{max}$ and $M(r_\mathrm{max})$.
As a quality check for a true maximum, we require that the circular velocity is smaller than the candidate maximum value everywhere within the extended region $f_\mathrm{check} r_\mathrm{max}$.
Here we use $f_\mathrm{check} = 1.5$.

The mass density profile of subhalos typically shows an uprise at large distances from their center due to the inclusion of host halo particles.
We can estimate a truncation radius for subhalos by first finding where the slope $\gamma$ of the enclosed density profile becomes shallow enough.
We take the maximum truncation radius defined by $\gamma(r_\mathrm{truncmax}) = -0.5$.
Using the enclosed average mass density has the advantage that it is much smoother than the local density profile.
From $r_\mathrm{truncmax}$ we then go inward in radius and find the minimum local density.
The location of the minimum local density is used as the truncation radius $r_\mathrm{trunc}$.
Assuming that the background density of the host halo at the location of the subhalo is given by $\rho(r_\mathrm{trunc})$, we can remove this background density and calculate the subhalo mass $M(r_\mathrm{trunc})$ and the circular velocity curve.

\bibliography{RDB_S}

\end{document}